\documentclass[iop]{emulateapj}
\usepackage{amsmath}
\usepackage{amsfonts}
\usepackage{amssymb}
\usepackage{booktabs}

\usepackage{graphicx}
\usepackage{color}
\usepackage{natbib}
\usepackage{needspace}

\newcommand{\E}[1]{\ensuremath{\langle #1 \rangle}}
\newcommand{\sigl}{\ensuremath{\sigma_{\!\mathcal{L}}}}
\def\Msun{\ensuremath{\mathrm{M_{\odot}}}}

\begin{document}

\submitted{Accepted for publication in \apjs}
\title{High-energy electromagnetic offline follow-up of LIGO-Virgo gravitational-wave binary coalescence candidate events}

\author{L.~Blackburn\altaffilmark{1,2}, M.~S.~Briggs\altaffilmark{3}, J.~Camp\altaffilmark{1},
    N.~Christensen\altaffilmark{4},
V.~Connaughton\altaffilmark{3}, P.~Jenke\altaffilmark{3},
    R.~A.~Remillard\altaffilmark{6}, \and J.~Veitch\altaffilmark{7}}
\altaffiltext{1}{NASA/Goddard Space Flight Center, Greenbelt, MD}
\altaffiltext{2}{University of Maryland/CRESST, College Park, MD}
\altaffiltext{3}{University of Alabama in Huntsville, Huntsville, AL}
\altaffiltext{4}{Carleton College, Northfield, MN}
\altaffiltext{5}{NASA/Marshall Space Flight Center, Huntsville, AL}
\altaffiltext{6}{Massachussetts Institute of Technology, Cambridge, MA}
\altaffiltext{7}{University of Birmingham, Birmingham, UK}

\begin{abstract}
We present two different search methods for electromagnetic counterparts to
gravitational-wave (GW) events from ground-based detectors using archival NASA
high-energy data from the Fermi-GBM and RXTE-ASM instruments. To demonstrate
the methods, we use a limited number of representative GW
background noise events produced by a search for binary neutron star
coalescence over the last two months of the LIGO-Virgo S6/VSR3 joint science
run. Time and sky location provided by the GW data trigger
a targeted search in the high-energy photon data. We use two custom pipelines:
one to search for prompt gamma-ray counterparts in GBM, and the other to search for
a variety of X-ray afterglow model signals in ASM. We measure the efficiency of
the joint pipelines to weak gamma-ray burst counterparts, and a family of model
X-ray afterglows. By requiring a detectable signal in either electromagnetic
instrument coincident with a GW event, we are able to reject a large majority of
GW candidates. This reduces the signal-to-noise of the loudest surviving GW
background event by around 15-20\%.
\end{abstract}

\section{Introduction}

\subsection{Initial and Advanced LIGO and Virgo detectors}

\begin{figure}
\begin{center}
    \hspace{.2in}
\includegraphics[width=3.1in]{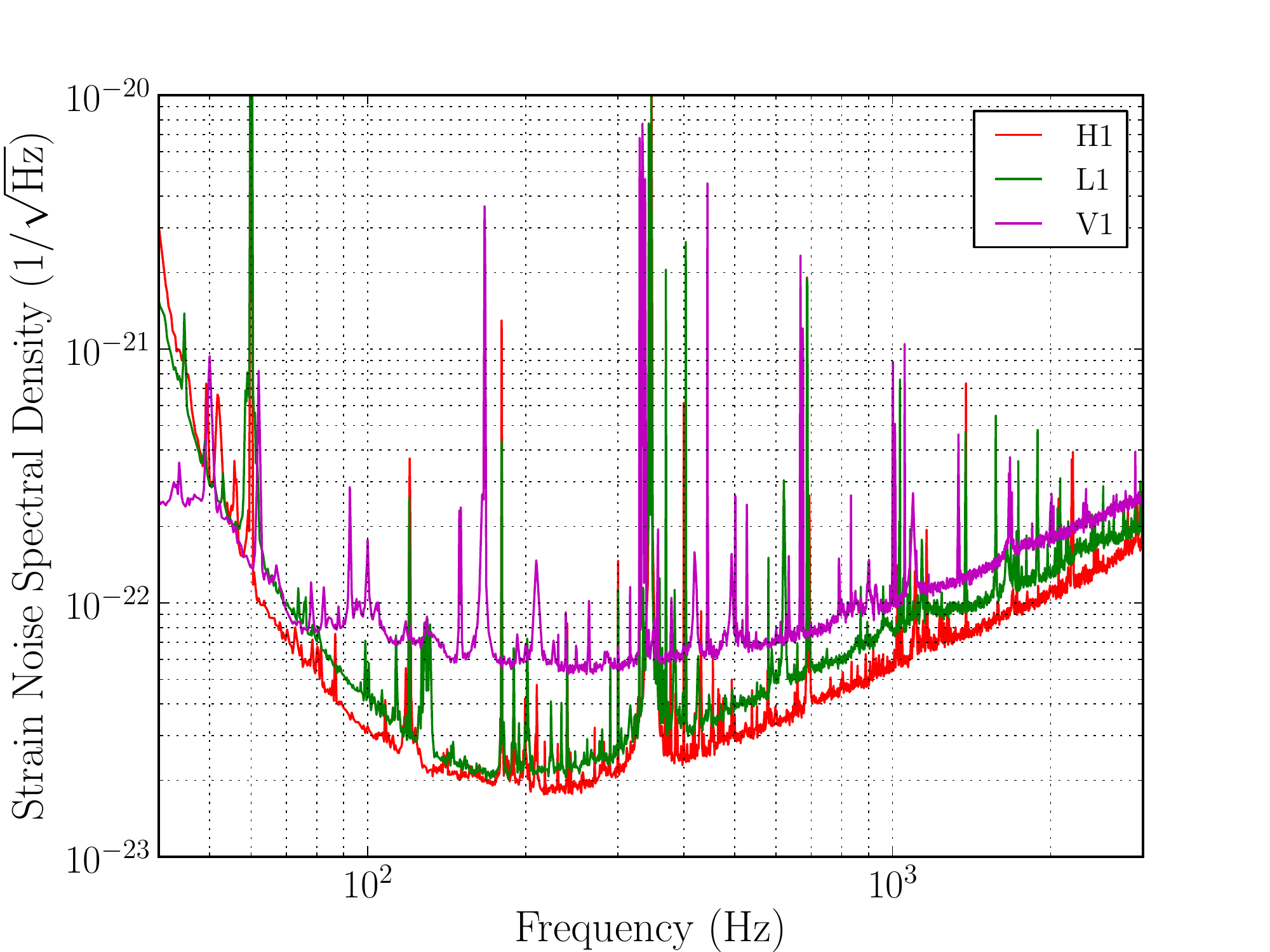}
\includegraphics[width=2.9in]{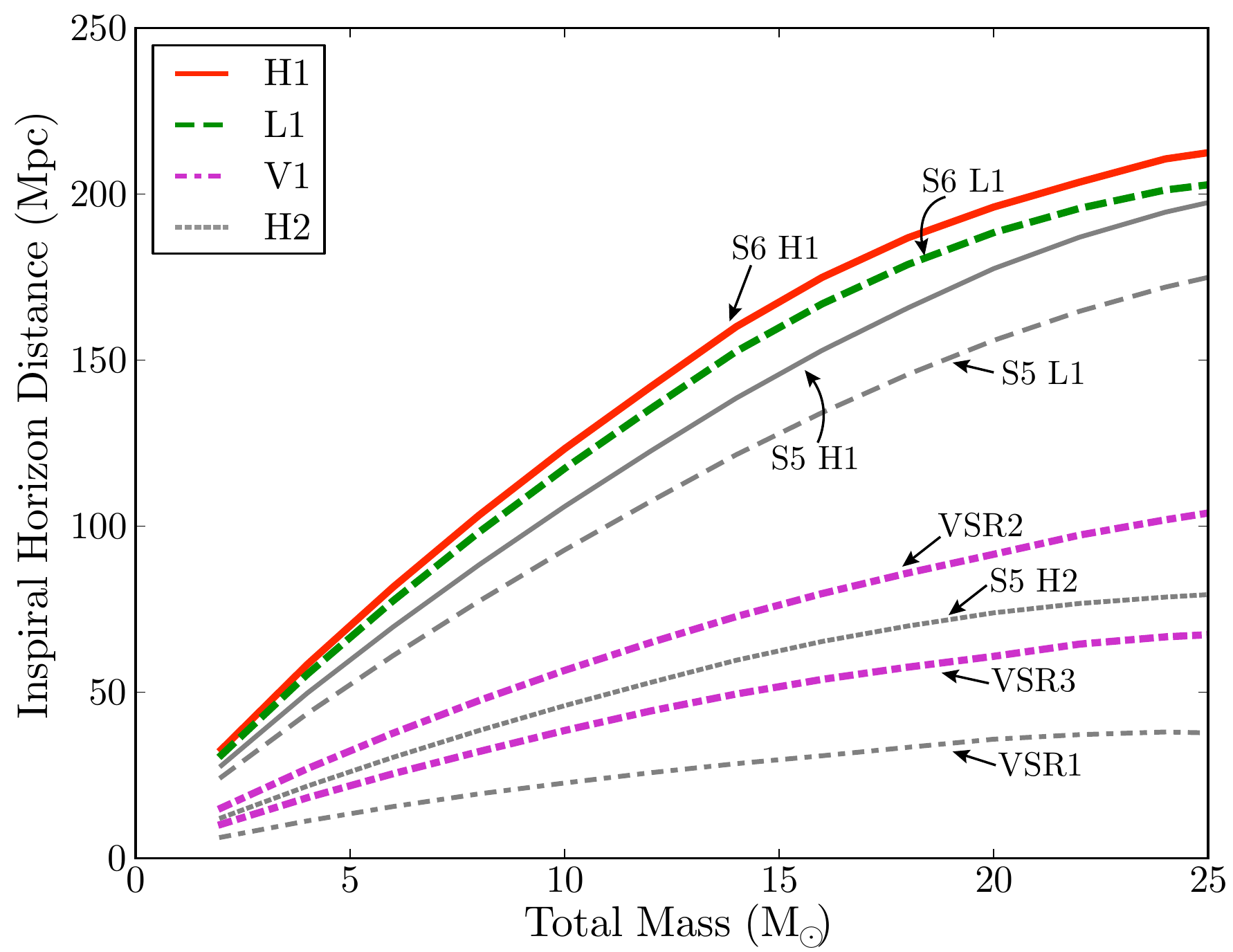} \caption{Typical detector
	strain noise spectral density for the LIGO S6 and Virgo VSR2+3 runs, as
	well as the best equal-mass binary coalescence horizon distance achieved
	for each run as a function of total system mass. The horizon distance is
	the distance at which an optimally oriented binary merger would produce an
	expected signal-to-noise of 8. Figures reproduced from
	\citep{Colaboration:2011nz}}
\label{fig:gwsensitivity}
\end{center}
\end{figure}

The initial LIGO \citep{Abbott:2007kv} and Virgo \citep{Accadia:2012zzb} gravitational-wave (GW) detectors took their last science
data between July 2009 and October 2010 before going offline for several years
for major upgrades to advanced detector configurations \citep{Harry:2010zz,17216}. 
For its sixth science run (S6),
LIGO consisted of two 4km baseline interferometric detectors - one instrument
H1 at Hanford, WA, and another instrument L1 at Livingston, LA. The Virgo 3km
interferometer V1 in Cascina, Italy took concurrent data during their second
(VSR2 covering July 2009 to Jan 2010) and third (VSR3 covering Aug 2010 to Oct
2010) science runs. Together, they represent the most sensitive worldwide
network of GW detectors to date. Figure \ref{fig:gwsensitivity}
shows typical strain sensitivity of each instrument as a function of frequency,
as well as the distance at which an optimally oriented merger of two compact
objects (black holes or neutron stars) would produce a nominal signal-to-noise
of eight in each detector.

Compact binary coalescence (CBC) is the most anticipated GW
source for first and second-generation ground-based detectors. These systems
are very strong emitters of gravitational waves, and we are confident of their
existence through the discovery of a handful of galactic NS/NS binary systems
where one object is a radio pulsar which is modulated by the orbit -- the most
famous of these systems is the Hulse-Taylor pulsar PSR1316+16
\citep{Weisberg:2004hi}.  The number and lifetime of these systems can be used
to obtain an estimate of about one NS/NS merger event per Mpc$^3$ per million
years \citep{Abadie:2010cf}, with up to two orders of magnitude uncertainty due
largely to our limited knowledge of the pulsar luminosity function and limited
statistics. The merger rate translates into an estimate of $\sim$0.02
detectable NS/NS merger events per year for the initial detector network, and
$\sim$40 per year for the advanced detectors once they reach design
sensitivity. NS/BH systems are also of interest for electromagnetic (EM)
follow-up and are stronger GW emitters.  However we have not
yet observed any NS/BH binary systems, and have generally poor knowledge of the
black hole mass distribution.

While no NS/NS or NS/BH binary coalescence events have been
detected during S6/VSR2+3 using GW data alone
\citep{Colaboration:2011nz}, it is conceivable that an EM
counterpart might allow us to resolve a rare event otherwise too weak to
distinguish from background in the first generation detectors. In the rapidly
approaching advanced detector era, searching for EM counterparts
can play an important role in increasing our confidence of detection for
otherwise marginal events and will provide astrophysical context for
GW detections.

Advanced LIGO is likely to begin taking its first science data in 2015, with
Virgo following a year later \citep{Aasi:2013wya}. As they reach design
sensitivity, the advanced detectors are expected to begin an era of regular
detection of GW events, making the search for EM counterparts triggered by GW
detections an enticing possibility. In this study, we demonstrate strategies
for searching high-energy archival EM data for counterparts using
representative GW background events from real data taken during the most recent
LIGO-Virgo science runs and high-energy EM survey data recorded at the same time. We
then measure EM detection efficiencies under various plausible emission models
and the probability of accidental coincidence given realistic GW background and
sky-localization accuracy.

\Needspace*{3\baselineskip}

\subsection{High-energy photon survey instruments}

The two instruments chosen for this EM follow-up study include the
Gamma-ray Burst Monitor (GBM) \citep{Meegan:2009qu} aboard the Fermi
spacecraft, and the soft X-ray All-Sky Monitor (ASM) \citep{Levine:1996du}
aboard RXTE. Both are particularly well suited to an offline follow-up of
GW events because of their large, regular coverage of the
entire sky (table \ref{tab:instruments}), and because they save a large amount of archival survey data which
allows for sensitive offline searches.


The high-energy sky itself offers the advantage of being relatively clean
compared to optical wavelengths, which is important for maintaining a low
probability of false-coincidence given the large GW sky-location uncertainty.
The use of offline survey data for follow-up has many practical differences
from other {\em triggered} searches in the EM spectrum for GW counterparts.
\citep{2012ApJS..203...28E,2014ApJS..211....7A,2012ApJ...759...22K}. The offline search does
not rely on rapid GW data analysis and continuous coordination with EM
observational facilities. An important consequence is that the offline coincidence
can generally be applied to many more events, allowing the follow-up of weaker, more
marginal candidates.

\begin{table}
	\begin{center}
	\begin{tabular}{ccccc}
		\toprule
		Instrument & Energy & FOV & Resolution & Cadence \\
		\midrule
		Fermi GBM & 8 keV -- 40 MeV & 65\% & $>$5$^\circ$ & 1.5 hr \\
		RXTE ASM & 1 keV -- 10 keV & 3\% & 0.1$^\circ$ & 1.5 hr \\
		\bottomrule
	\end{tabular}
    \caption{High-energy photon survey instruments used to search for EM counterparts. Field-of-View (FOV) is represented as a percentage of the entire sky.}
\label{tab:instruments}
\end{center}
\end{table}

\Needspace*{3\baselineskip}
\subsection{Short GRBs and afterglows as counterparts to GWs}

Gamma-ray bursts (GRB) are flashes of gamma rays observed approximately once
per day. Their isotropic distribution in the sky was the first evidence of an
extra-galactic origin, and indicated that they were extremely energetic events.
The duration of prompt gamma-ray emission shows a bi-modal distribution which
naturally groups GRBs into two categories \citep{Kouveliotou:1993yx}. Long GRBs
emit 90\% of their prompt radiation on timescales longer than 2 seconds.  They
have been associated with the collapse of rapidly rotating massive stars
\citep{Hjorth:2011zx}.  Short GRBs (sGRB) with prompt emission $<$2s and a generally
harder spectrum are thought to arise from the merger of two neutron stars, or a
neutron star and black hole \citep{Nakar:2007yr}. It is this favored progenitor
model which makes short GRBs and associated afterglow emission a promising
counterpart to GW observations.

The Swift satellite has revolutionized our understanding of short GRBs over the
last several years by the rapid observation of X-ray afterglows, providing the
first localization, host identification, and red-shift information. The beaming
angle for short GRBs is highly uncertain, although limited observations of jet
breaks in some afterglows imply half-opening angles of $\theta_j \sim$ 3--14$^\circ$
\citep{Liang:2007rn,Fong:2012aq}. The absence of an observable jet break
sets a lower limit on the opening angle which is generally weak (due to limits
in sensitivity), though in the case of GRB 050724A, late-time Chandra
observations were able to constrain $\theta_j \gtrsim 25^\circ$
\citep{Grupe:2006uc}. The observed spatial density of short GRBs and limits on
beaming angle result in a NS/NS merger event rate roughly consistent with that
derived from galactic binary pulsar measurements.

Although the beaming factor
of $\sim$$\theta_j^2/2$ means most merger events seen by the advanced GW
detectors will not be accompanied by a gamma-ray burst, this is somewhat
compensated by the fact that the ones that are beamed toward us have stronger
GW emission. Current estimates for coincident GW-sGRB
observation for advanced LIGO-Virgo are a few per year assuming a NS/NS
progenitor model \citep{Metzger:2011bv,Coward:2012gn}.  The rate increases by a
factor of 8 if all observed short GRBs are instead due to NS/BH (10 $M_\odot$)
mergers which are detectable in gravitational waves to twice the distance.
Coincident rates for the initial detectors go down by a factor a thousand due
to their factor of ten worse sensitivity.

About 80\% of short GRBs seen by Swift are accompanied by some kind of X-ray
afterglow \citep{Gehrels:2013xd}, which are often observable for about a day.
The observed X-ray afterglows are thought to occur from synchrotron emission at
the shock front where the outgoing jet meets the local inter-sellar medium. Simulations show
that such afterglow emission becomes very weak off-axis, with little
possibility of detection at twice the opening angle \citep{vanEerten:2011zu}.
Searches for orphan afterglow signals, without the presence of detected prompt
GRB emission, are quite difficult to confirm due to various sources of
transient background.  However with incomplete coverage of the gamma-ray sky,
it is quite possible that the first GW-EM association will be with an afterglow
signal.

In addition to the jet-driven burst and afterglow, other EM emission associated
with a compact merger can be a promising channel for GW-EM coincidence,
particularly if the EM radiation is less-beamed or even isotropic. A few short
GRBs ($\sim$10\%) have shown clear evidence of high-energy flares which preceed
the primary burst by 1--10 seconds, and possibly up to 100s
\citep{Troja:2010zm}. The precursors can be interpreted as evidence of some
activity during or before merger, such as the resonant shattering of NS crusts
\citep{Tsang:2011ad}, which depends on the NS equation-of-state and could
radiate isotropically. A class of short GRBs ($\sim$25\%) also contain a period
of extended soft X-ray emission on a timescale of $\sim$10--100s after the
initial spike, which is difficult to explain with the standard jet scenario. It
has been proposed that the extended emission could arise from a relativistic
wind powered by a short-lived rapidly-rotating protomagnetar star surviving
post-merger
\citep{2008MNRAS.385.1455M,2013ApJ...763L..22Z,Gao:2013rd,Nakamura:2013hda},
which could also be considerably more isotropic than the jet afterglow. Since
short GRBs themselves may be subject to a large beaming factor, the more
numerous nearby NS/NS mergers seen in gravitational waves will provide a unique
opportunity to search for these potentially weak exotic EM phenomena, and
isolate them from jet behavior.

Detection of the characteristic signature of a compact binary coalescence in
gravitational waves prior to an associated sGRB will provide unambiguous
support for the compact merger progenitor model. Along with this critical piece
of the sGRB puzzle, gravitational waves provide a largely complimentary set of
information: component masses (whether black holes are involved) and spins,
system inclination, luminosity distance; while the EM counterpart
provides information about EM energetics, a precise location, local and host
environment, and red-shift. This makes joint GW-EM detections particularly
valuable and a key goal in GW astronomy.

\subsection{Methodology of GW-EM coincidence search}

GW-EM coincidence can be approached from a variety of angles. In this
analysis, we use a collection of all-sky, all-time GW events to
conduct a targeted search in high-energy EM archival survey data. Another
strategy demands real-time analysis and localization of the GW data
\citep{2012A&A...541A.155A,Cannon:2011vi,Singer:2014qca}, suitable for targeted
observation with narrow-field instruments with the goal of catching a
short-lived afterglow signal. This rapid EM follow-up approach was succesfully
tested in initial LIGO/Virgo for a handful of online GW events using both a
collection of ground-based optical telescopes \citep{2014ApJS..211....7A} and
Swift-XRT \citep{2012ApJS..203...28E}. The online strategy places heavy demands
on both the rapid processing and interpretation of GW data, as well as the
capability of pointed telescopes to effectively cover the large GW-determined
sky location \citep{Abbott:2011ys,2012ApJ...759...22K,Singer:2013xha}. In
return they provide the deepest observations, with sufficient sensitivity to
observe, for example, the faint but promising optical/IR kilonova merger
counterpart \citep{Metzger:2011bv,Kasliwal:2013yqa}.

Known EM transients can also trigger specialized searches in the GW data
itself, where the precise time, sky-location, and potentially model information
from the EM event may greatly constrain the space of GW signals to look for.
The reduced search can be performed at a lower threshold in the GW data due to
both computational and statistical (false-alarm probability) considerations
\citep{Was:2012zq}. In the case of known GRBs, such searches have been used to
place minimum distance limits dependent on GW emission model
\citep{Briggs:2012ce,Aasi:2014iia}, and for two short GRBs have been able to
uniquely exlude the possibility of a binary merger progenitor from a plausible
nearby host \citep{Abbott:2007rh,Abadie:2012bz}.

For this anaysis, the offline coincidence between GW and EM
observation is done in time and sky location, providing an ability to exclude
non-coincident background from either search. Timing for a GW
signal is very good, and the time of coalescence is accurate to 
milliseconds. The coincidence search window in time is
determined by the expected delay between coalescence and the emission of
EM radiation, which ranges from seconds for prompt gamma-rays to
hours for afterglow emission.  Sky localization using gravitational waves is
based primarily on triangulation across a light-travel baseline of 10--30 ms
\citep{Fairhurst:2010is}, but can be improved by including the signal amplitude
and phase \citep{2014PhRvD..89h4060S,Singer:2014qca}. For the initial detector network, sky
position for a merger event can be determined to several tens of square degrees
\citep{Fairhurst:2009tc}.

\begin{figure}
	\begin{center}
        \includegraphics[width=3in]{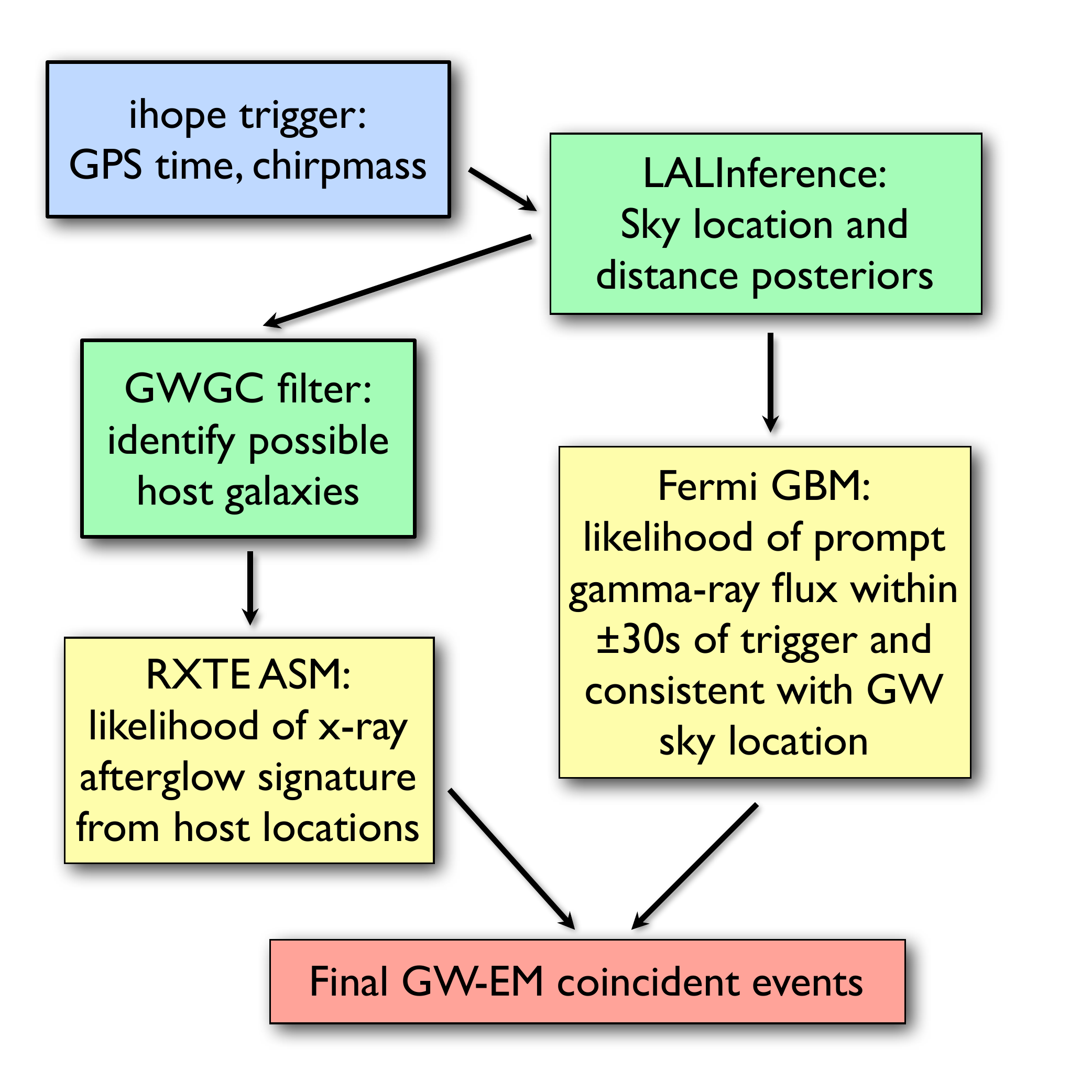} \caption{Flowchart
            for GW-EM coincidence analysis. We begin with GW
            compact binary coalescence events detected with a standard
            matched-filter analysis (ihope) \citep{Babak:2012zx}. A coherent Bayesian follow-up using the LALInference package \citep{Veitch:2014wba} estimates
            distance and sky location parameters using data from multiple
            detectors. From this, we obtain a list of possible nearby host
            galaxies from the galaxy catalog (GWGC) and their relative probabilities. We search nearby GBM
            data for prompt gamma-ray bursts consistent with the GW skymap, as
            well as ASM light-curve data at the positions of potential host
            galaxies for a family of parameterized afterglow signals beginning
            at the coalescence time. From this we gather a filtered list of
            GW-EM candidate events.}
		\label{fig:flowchart}
	\end{center}
\end{figure}

Figure \ref{fig:flowchart} shows an outline of the GW-EM coincidence pipeline.
We begin with GW events found by a standard matched-filter
analysis for coincident CBC signals across two or all three of the GW detectors
\citep{Colaboration:2011nz}. The original results of this analysis identified a
loud signal that had been injected into the data as a blind test of the
pipeline (as seen in figure \ref{fig:ihopedistribution}), but did not find any
other outliers that stood out above the coincident background of the GW
instruments. The matched filter analysis gives some information about sky
location (from timing) and distance (from amplitude), but better parameters are
obtained through a coherent Bayesian follow-up \citep{Veitch:2009hd,Veitch:2014wba} which
calculates the posterior probability of all possible physical signals (with
varying component masses, location, distance, inclination, etc) using their
coherent overlap with the data from all detectors. This is particularly useful
for the relatively common case of a signal detected in H1 and L1, but too weak
to be detected in Virgo, thus missing V1 timing information. The Bayesian
analysis is still able to use the Virgo data to exclude regions of the sky
where, for example, Virgo is particularly sensitive relative to the LIGO
instruments.

The sky location and distance estimates can be further refined by assuming that
compact mergers arise from a known galaxy, for which we have reasonably
complete catalogs out to the LIGO horizon \citep{White:2011qf}. Each galaxy in
the catalog can be assigned a host probability which is proportional to its
blue-light luminosity (as a proxy for its mass, which we assume is proportional
to the rate of compact binary mergers), as well as its overlap with the
GW-derived posterior distributions in distance and sky-location. This reduces
the search from several hundred square degrees to a handful of essentially
point-source locations (given the angular resolution of ASM), at the risk of
the true host being absent from the catalog.

The offline search for an EM counterpart from GBM or ASM is then
triggered by the time of coalescence, and the GW-derived skyamp or the list of
possible host galaxies with corresponding probabilities. We search over the
parameter space of expected EM signals, and set nominal thresholds
above which we consider GW-EM candidate coincident events.

\Needspace*{5\baselineskip}
\section{LIGO-Virgo event generation}

\subsection{Matched filter CBC search}

As two compact objects orbit, they lose orbital energy and angular momentum to
gravitational waves. The orbit responds by shrinking and the orbital period decreases, causing
even more rapid loss of orbital energy to gravitational radiation. This process
ultimately leads in a runaway process to merger. At any moment during the
inspiral, the GW emission is well characterized by high-order
analytic post-Newtonian approximations to general-relativity. By following the
orbit adiabadically through the detector bandwidth, a characteristic chirp
waveform of increasing amplitude is produced that sweeps through the sensitive
band ($\sim$40--2000~Hz) of ground-based GW detectors.

LIGO-Virgo GW data from S6/VSR2+3 has been searched for compact
binary coalescence events with total system mass $<$ 25 M$_\odot$, and the
results have been reported in \cite{Colaboration:2011nz}. The search made use
of matched filtering, correlating the GW data in each detector
against a bank of theoretical templates composed of model inspiral
chirp signals for non-spinning systems with various component masses and
orientations. Matched filtering describes the optimal linear filter for
maximizing signal-to-noise in the presence of stationary Gaussian noise. For
the filter corresponding to a frequency domain signal $\tilde{h}(f)$, the
matched filter produces an expected single-detector signal-to-noise ratio (SNR) of,
\begin{equation}
	\rho^2 = \int_0^\infty{\frac{4|\tilde{h}(f)|^2}{S_n(f)}df}
\end{equation}
where $S_n(f)$ is the one-sided power spectral density of the stationary noise.

Due to the presence of non-Gaussian noise in the detector data, a $\chi^2$
consistency test \citep{Allen:2004gu} is applied to events which are initially
identified as local maxima in matched-filter SNR. A re-weighted SNR
$\hat{\rho}$ is obtained by downgrading events using an empirically-determined
relationship that takes the level of inconsistency with the model signal into
account \citep{Babak:2012zx},

\begin{equation}\label{eqn:new_snr} \hat{\rho} = \begin{cases} {\displaystyle
\frac{\rho}{[(1+(\chi^2_r)^3)/2]^{1/6}}} & \mbox{for } \chi^2_r > 1, \\ \rho &
\mbox{for } \chi^2_r \leq 1.  \end{cases}  \end{equation}

\begin{figure}[t]
\begin{center}
\includegraphics[width=3in]{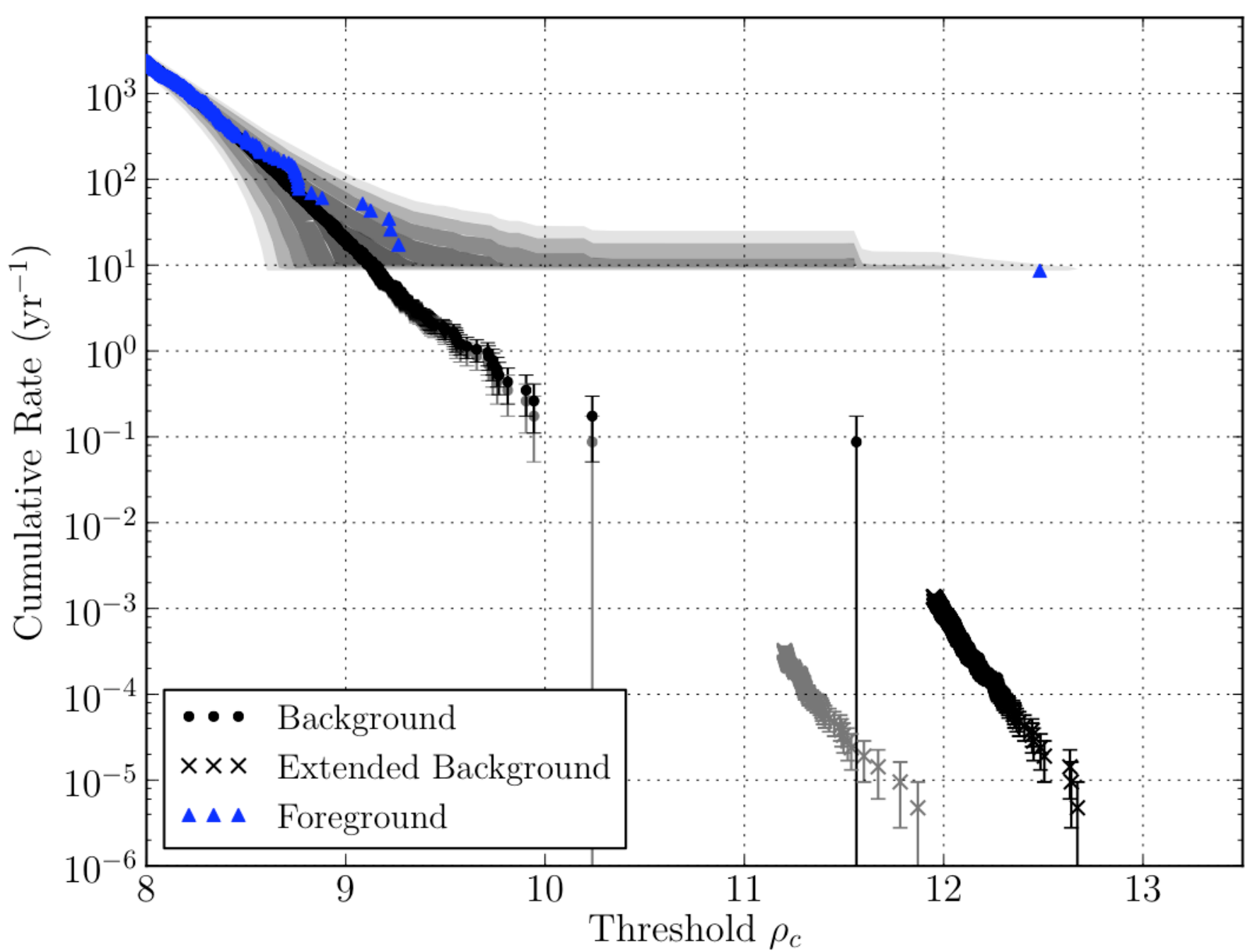} \caption{The
	S6/VSR2-3 science run contained a simulated NS/BH merger signal that was
	injected into the data without knowledge of the analysis teams. This plot
	from \citep{Colaboration:2011nz} shows the cumulative rate of events
	coincident in the H1 and L1 detectors with chirp mass $3.48 \le
	\mathcal{M}_c/\Msun < 7.40$ as seen by the matched filter pipeline in four
	months of data around the simulation. Chirp mass is a function of the individual
    masses: $\mathcal{M}_c=(m_1 m_2)^{3/5}/(m_1+m_2)^{1/5}$, and is the dominant
    mass parameter driving binary evolution during inspiral. The injection, shown by the blue
	triangle with ranking statistic $\rho_c$ = 12.5, is clearly resolved
	against the expected background distributions derived from time-shift
	analysis, both with (black) and without (gray) single-detector
	contributions from the simulated event in one of the detectors (coincident
	with time-shifted noise in another). Additional coincidence with an
	EM counterpart can help resolve any astrophysical events
	present in the data in regions with otherwise large background ($\rho_c \sim$9.5).}
\label{fig:ihopedistribution}
\end{center}
\end{figure}

Events from two or more detectors are searched for coincidence in
time and mass parameters. Coincident events are ranked by the quadrature sum,
$\rho_c$, of their re-weighted signal-to-noise. This {\em combined SNR} for each event is
compared against an estimated background distribution derived from running the
same coincidence analysis many times while applying unphysical relative
time-shifts between the detector data (as in figure
\ref{fig:ihopedistribution}). When the event has a $\rho_c$ value
above a predetermined threshold corresponding to some low false-alarm rate (e.g. one false alarm from background per
several thousdand years), the
event is considered a GW candidate. Background estimation is performed
separately for various two-week epochs, detector combinations, and mass bins in
order to isolate configurations with varying background levels.

The S6/VSR2+3 analysis identified one outlier event with an estimated
false-alarm-rate of $\sim$1/7000 yr$^{-1}$. The event was ultimately revealed
as a blind simulated signal injected in the data, which served as an end-to-end
test of the detection strategy.\footnote{details at
\url{http://www.ligo.org/science/GW100916/}} Otherwise, no events stood out
clearly from the expected background.

\subsection{Bayesian parameter estimation}


Bayesian integration is a natural way to determine whether or not a coherent
GW signal is present in the data from a network of instruments,
and if so, what are the likely parameters.  While the coincident matched-filter
analysis described in the previous section imposes some consistency on time and
mass, a coherent analysis requires from the start a common physical
gravitational waveform projected onto each instrument, making it particularly
useful for a network of detectors.

If we let $\mathcal{H}_1$ represent the hypothesis that a CBC signal is present
in the data, and let $\mathcal{H}_0$ represent only noise, the likelihood ratio
provides the optimal statistic to distinguish the two,
\begin{equation}
	\label{eqn:inspnestlr}
	\Lambda = \frac{P(d|\mathcal{H}_1, I)}{P(d|\mathcal{H}_0, I)}
\end{equation}
where $I$ represents any prior information we have about the system. Given that
our signal hypothesis $\mathcal{H}_1$ represents a large population of possible
signals with different waveforms determined by their set of binary parameters
$\vec{\theta}$, proper evaluation of the likelihoood involves marginalizing over
the parameter space,
\begin{equation}
	\label{eqn:inspmarg}
	P(d|\mathcal{H}_1, I) = \int{P(\vec{\theta}|\mathcal{H}_1, I) P(d|\mathcal{H}_1, \vec{\theta}, I) d\vec{\theta}}
\end{equation}
where $P(\vec{\theta})$ represents the prior probability distribution of
parameters $\vec{\theta}$ in our signal population -- for example it may
reflect that we expect systems to be uniformly distributed in volume with
random orientations with respect to the detectors.

In addition to the question of whether or not a signal is present, we are also
interested in determining the physical parameters of the binary system that
are implied by a set of
data. This is represented by the posterior probability distribution over the
parameters $\vec{\theta}$,
\begin{equation}
	\label{eqn:inspnestposterior}
	P(\vec{\theta}|d,\mathcal{H}_1,I) = \frac{P(\vec{\theta}|\mathcal{H}_1,I)P(d|\vec{\theta},\mathcal{H}_1, I)}{P(d|\mathcal{H}_1,I)}
\end{equation}
If we are interested in the probability distribution only over a subset
$\vec{\theta}_A$ of the parameters, where $\vec{\theta} \equiv \{\vec{\theta}_A,
\vec{\theta}_B\}$, we marginalize over those that remain,
\begin{equation}
	P(\vec{\theta}_A|d,\mathcal{H}_1, I) = \int{P(\vec{\theta}|d,\mathcal{H}_1, I) d\vec{\theta}_B}
\end{equation}

Nested sampling \citep{2004AIPC..735..395S} is a computationally efficient alternative to MCMC techniques \citep{Christensen:2001cr},
both of which are designed to intelligently sample and integrate the
probability distribution of a high-dimensional parameter space of possible
signals. For the case of CBC, we assume the spin of each neutron
star is small and there are nine relevant physical parameters: two
for the component masses, sky location, and orientation, as well as distance,
time, and phase. We apply a Bayesian coherent follow-up based on nested sampling
\citep{Veitch:2009hd} to a large number of coincident events identified from
the matched filter analysis. The output of the nested sampling routine is a set
of sample vectors which trace the estimated 9-dimensional posterior probability
distribution of system parameters (equation \ref{eqn:inspnestposterior}), as
well as the integrated likelihood obtained by summing over them all (equation~\ref{eqn:inspmarg}).

\subsection{Galaxy targeting}

Extra-galactic events detectable by the initial LIGO-Virgo network are close
enough so that a galaxy catalog can be used to identify probable hosts. While
this is only marginally useful for a search using GW data alone due to the poor
angular resolution of current GW detector networks, it becomes quite important
for EM follow-ups where the EM sky resolution is significantly
better \citep{2013ApJ...767..124N,Hanna:2013yda}. Inclusion of a galaxy catalog
for this offline EM search greatly reduces the computational cost and
false-positive rate of the ASM analysis by reducing an area of $\sim$150 square
degrees to just tens of individual points.

The Gravitational-Wave Galaxy Catalog (GWGC) \citep{White:2011qf} was designed
specifically to aid in GW searches with the initial detector network. It
contains distance, type, location, geometry, and blue-light luminosity
information for $\sim$50k galaxies within 100 Mpc, taken from the HyperLEDA
database and other sources. To choose probable host galaxies from the catalog,
we compare their distance and sky location against the probability distribution
derived from the Bayesian analysis on GW data. Galaxies which occur in regions
of high probability are considered for EM follow-up. At advanced LIGO/Virgo
distances, galaxy catalogs suffer from incompleteness, with only the brighter
galaxies making the flux limits of surveys, and deep surveys often covering
only fractions of the sky. \cite{Hanna:2013yda} outline some of the
considerations necessary to effectively use incomplete catalogs in that regime.
For this initial LIGO/Virgo study, we assume the GWGC is 100\% complete out to
the distance of detectable NS/NS mergers.

The GW posterior probability distribution must be estimated from the discrete
sampling provided by the nested sampling procedure. We estimate the density of
posterior samples at each galaxy location and distance using Gaussian kernel
density estimation (KDE). We treat distance and sky location independently due
to limited sampling of the posterior, typically thousands of points.

\begin{figure}
\begin{center}
\includegraphics[width=3.4in]{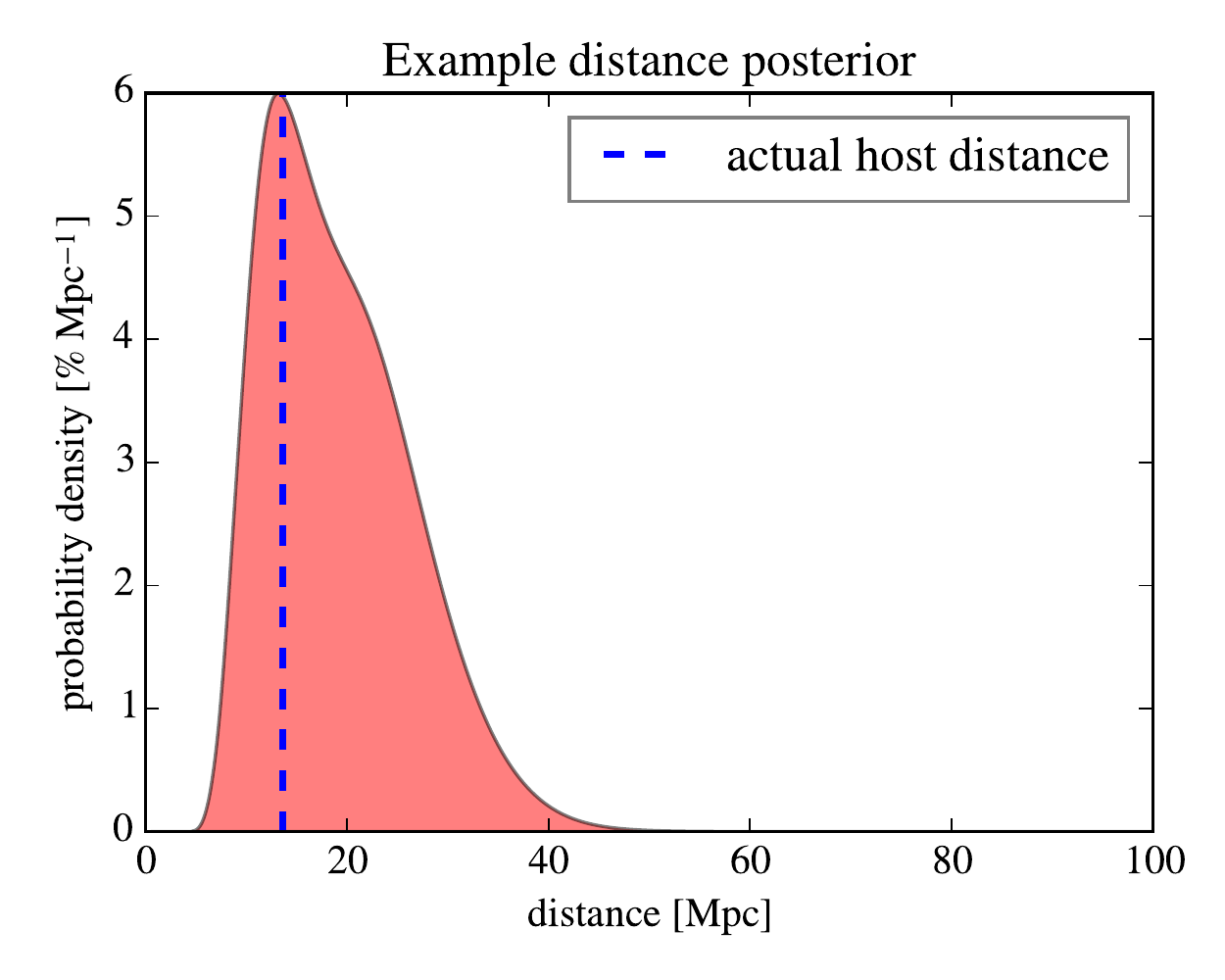}
\includegraphics[width=3.4in]{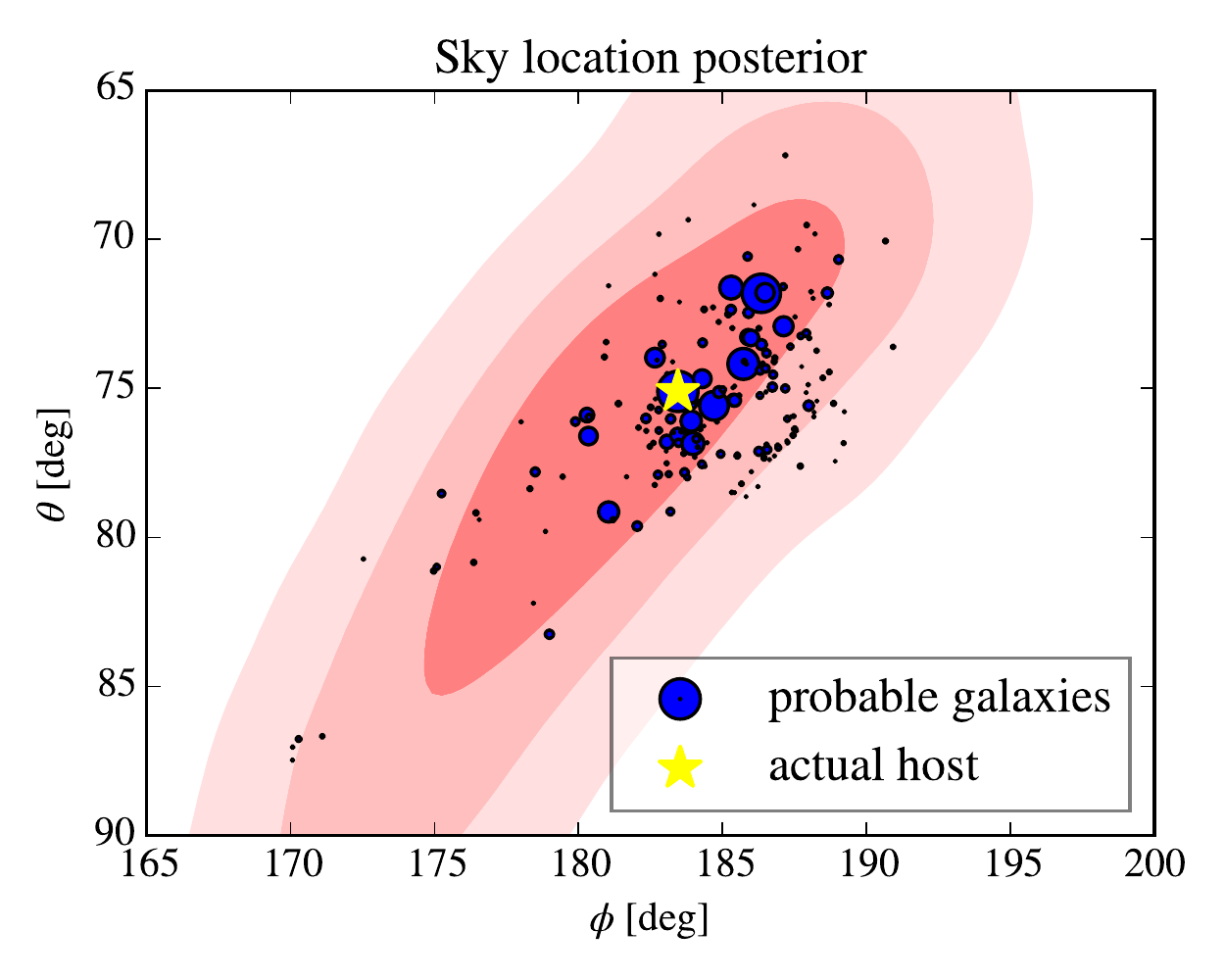}
\caption{ Distance (top) and
	sky position (bottom, with spherical coordinates $\theta$, $\phi$ in radians) posterior distributions derived from the Bayesian
	follow-up of GW data are used to filter a galaxy catalog to
	check for likely nearby hosts (bottom). The area of the weighted galaxies
	corresponds to their relative probability of being the host. It is directly
	proportional to the blue-light luminosity (as a proxy for mass, or merger
	rate), and the estimated posterior probability distribution at the
	corresponding distance and sky location. For this simulation of a
	misaligned NS/NS merger in the Virgo cluster, the maximum probability galaxy
	corresponed to the true host.}
\label{fig:gwgcfiltering}
\end{center}
\end{figure}

For a galaxy with blue-light luminosity $L_{10}$ ($10^{10} L_\odot$) at
distance $r < 100$ Mpc and location $\Omega$, the estimated distribution from
the $N$ posterior samples is,
\begin{gather}
    P_\mathrm{GW}(r) = \frac{1}{N} \sum_i^N{\frac{1}{r^2}w(r_i-r, \sigma_r = \text{10--20\%})} \\
    \label{eqn:pomega}
    P_\mathrm{GW}(\Omega) = \frac{1}{N} \sum_i^N{w(\vec{x}_i-\vec{x},\, \sigma_\Omega=3^\circ)} \\
	\label{eqn:gwgcprobability}
	P(L_{10}, r, \Omega) = L_{10} P_\mathrm{GW}(r) P_\mathrm{GW}(\Omega)
\end{gather}
where $w(\mu,\sigma)$ represents the 1D and 2D Gaussian kernels, normalized
to give $P = 1$ in the case of a uniform posterior. Figure
\ref{fig:gwgcfiltering} shows an example of the kernel density estimation
applied to the distance parameter, as well as a sample of matching galaxies,
weighted by all three factors in equation \ref{eqn:gwgcprobability}, against
the GW skymap. The distance KDE bandwidth is chosen according to the estimated
individual galaxy distance errors, which are around 10--20\% depending on the
original source catalog. The 3$^\circ$ sky bandwidth is chosen to balance
statistical errors (resulting Poisson errors are at the level of a few~\%) with
resolution of the GW network.  Finally, we have undone the $r^2$ distance prior
from the assumption of homogeneously distributed sources during the Bayesian
integration as the galaxies themselves already include this volume effect.

\section{Gamma-ray counterparts with Fermi GBM}
\subsection{GBM instrument and data products}

The Gamma-ray Burst Monitor (GBM) \citep{Meegan:2009qu} aboard the Fermi
spacecraft measures photon rates from 8 keV--40 MeV. The instrument consists of
12 semi-directional NaI scintillation detectors and 2 BGO scintillation
detectors which cover the entire sky not occluded by the Earth (about 65\%).
The lower-energy NaI detectors have an approximately $\cos{\theta}$ response
relative to angle of incidence, and relative rates across detectors are used to
reconstruct the source location to a few degrees.  The BGO detectors are much
less directional, and can be used to detect and resolve the higher energy
spectrum above $\sim$200 keV.

GBM produces on-board triggers for gamma-ray burst events by looking for
multi-detector rate excess over background across various energy bands and
timescales. In the case of a trigger, individual photon information is sent to
the ground and the event is publicly reported. Those events which have been
confirmed as GRBs have already been studied in coincidence with LIGO-Virgo data
\citep{2012ApJ...760...12A}. In addition to the triggered events, survey data
is available which records binned photon counts over all time. For this offline
analysis of short transients, we use the CTIME (time-resolved) daily data, which is binned at
0.256s over 8 energy channels for each detector. A new GBM data product for
continuous time-tagged events
(CTTE) was introduced in 2012 that provides complete individual photon
information with 2$\mu$s and 128 energy channel resolution. This data product
was not available for un-triggered times during the initial LIGO-Virgo science
runs, but will enhance offline sensitivity to short bursts when the advanced GW
detectors begin operation.

\subsection{Coherent analysis of GBM data}

In this section, we develop a procedure to coherently search GBM detector data
for modeled events \citep{Blackburn:2013ina}. The idea is that by processing
multiple detector data coherently, we can obtain a greater sensitivity than
when considering one detector at a time. Greater computational resources
available offline (vs.  on-board) also allow for more careful background
estimation to be done.  For this analysis, we can relax to some extent the
strict 2-detector coincidence requirement used to veto spurious events on-board
as the GW trigger means much less time and sky area is
considered. These advantages help to balance out the coarse time resolution of
CTIME data, which reduces offline sensitivity to very short bursts prior to
2013.

\Needspace*{2\baselineskip}
Each detector is subject to a substantial time-varying background from bright
high-energy sources that come in and out of the wide field of view, as well as
location-dependent particle and Earth atmospheric effects. This background must
be estimated and subtracted out to look for any prompt excess.  In this
analysis where we are interested in the background estimate for a short
foreground interval [$-T/2$, $+T/2$] where $T \sim$1s, we estimate the
background using a polynomial fit to local data from [$-10T$, $+10T$] (minimum
$\pm5s$), excluding time [$-3T/2$, $+5T/2$] around the foreground interval to
avoid bias from an on-source excess. The polynomial degree is determined by the
interval length to account for more complicated background variability over
longer intervals. It ranges from 2 (minimum) to 1+$0.5\log_2 T$. A separate fit
is done for each detector/channel combination, and determines the background
rate estimation over the foreground interval, as well as its systematic
uncertainty from fitting error. Data from channels with poor fits (large
$\chi^2$) are excluded from the analysis.

High-energy cosmic rays striking a NaI crystal can result in long-lived
phosphorescent light emission. The detector may interpret this is a rapid
series of events, creating a short-lived jump in rates for one or multiple
channels, and severely distorting the background fit if not accounted for.
They are identified with a simple procedure that scans for rapid 1-bin spikes
in the background interval. The affected bins are removed from the background
fitting. Cosmic rays affecting the foreground interval are handled differently,
as described in section \ref{sec:selectioncuts}.

A likelihood ratio combines information about sources and noise into a single
variable. It is defined as the probability of measuring the observed data, $d$,
in the presence of a particular true signal $H_1$ (with some source amplitude
$s > 0$) divided by the probability of measuring the observed data in noise
alone $H_0$ ($s = 0$),  \begin{equation} \Lambda(d) = \frac{P(d | H_1)}{P(d |
H_0)}.  \end{equation} When signal parameters such as light-curve, spectrum,
amplitude $s$ and sky-location $\alpha, \delta$ are unknown, one can either
marginalize over the unknown parameters, or take the maximum likelihood over
the range to obtain best-fit values.

For binned, uncorrelated Gaussian noise, \begin{align} \label{eqn:lrnumerator}
    P(d_i|H_1) &= \prod_i{\frac{1}{\sqrt{2\pi}\sigma_{d_i}}
        \exp\left(-\frac{(\tilde{d_i}-r_is)^2}{2\sigma_{d_i}^2}\right)} \\
    \label{eqn:lrdenominator} P(d_i|H_0) &=
    \prod_i{\frac{1}{\sqrt{2\pi}\sigma_{n_i}}
        \exp\left(-\frac{\tilde{d_i}^2}{2\sigma_{n_i}^2}\right)} \end{align}
where we have used $\tilde{d_i} = d_i - \E{n_i}$ to represent the
background-subtracted measurements in each detector-time-energy bin,
$\sigma_{n_i}$ and $\sigma_{d_i}$ for the standard deviation of the background
and expected data (background+signal), $r_i$ for the
location/spectrum-dependent instrumental response, and $s$ for the intrinsic
source amplitude at the Earth. Maximizing the likelihood ratio is the same as
maximizing the log-likelihood ratio $\mathcal{L} = \ln{\Lambda}$,
\begin{equation} \label{eqn:loglikelihood} \mathcal{L} =
    \sum_i{\left[\ln{\frac{\sigma_{n_i}}{\sigma_{d_i}}}
            +\frac{\tilde{d_i}^2}{2\sigma_{n_i}^2}
            -\frac{(\tilde{d_i}-r_is)^2}{2\sigma_{d_i}^2}\right]}
\end{equation}

The dependence of the response factors $r_i$ on sky location is complicated, so the
likelihood ratio is calculated over a sample grid of all possible locations.
Assuming a single location, the remaining free parameter is the source
amplitude $s$. The variance in the background-subtracted detector data includes
both background and source contributions, \begin{equation} \label{eqn:sigmadi}
    \sigma_{d_i}^2 = \sigma_{n_i}^2 + r_is + \sigma_{r_i}^2s^2 \quad (s \ge 0)
\end{equation} with $\sigma_{r_i}^2$ representing Gaussian-modeled systematic
uncertainty in the instrumental response. Source terms are only included for
physical $s \ge 0$, else their contribution is zero. The background contributes
Poisson error, as well as any systematic variance $\sigma_{b_i}^2$ from poor
background fitting which is also assumed to be Gaussian, \begin{equation}
    \label{eqn:sigmani} \sigma_{n_i}^2 = \E{n_i} + \sigma_{b_i}^2.
\end{equation} We find $s_\mathrm{best}$ which maximizes $\mathcal{L}$ by
setting the derivative $d\mathcal{L}/ds$ to zero using Newton's method.

To consider all possible source amplitudes we need to integrate the likelihood
$P(d|s)$ (equation \ref{eqn:lrnumerator}) over a prior on $s$, \begin{equation}
    \label{eqn:marginalization} P(d) = \int{P(d|s)P(s)ds} \end{equation} For a
given set of detector data $d$, the likelihood $P(d|s)$ over $s$ is almost the
product of individual Gaussian distributions (not quite Gaussian because
$\sigma_d$ depends on $s$). The product of Gaussian distributions with mean
values $\mu_i$ and standard deviations $\sigma_i$ is itself Gaussian with mean
and variance, \begin{equation} \mu_\mathrm{prod} =
    \frac{\sum{\mu_i/\sigma^2_i}}{\sum{1/\sigma^2_i}}, \quad
    \sigma^2_\mathrm{prod} = \frac{1}{\sum{1/\sigma^2_i}} \end{equation} In
this case, $\mu_i = \tilde{d_i}/r_i$ and $\sigma_i = \sigma_{d_i} / r_i$.  We
estimate the variance of $\mathcal{L}$ over $s$ as, \begin{equation} \sigl^2 =
    \frac{1}{\sum{r_i^2/\sigma_{d_i}^2}}, \quad \sigma^2_{d_i} \text{ evaluated
        at } s_\mathrm{best} \end{equation} and choose a scale-free prior with
fixed $\beta = 1$, so that our choice of form for an amplitude prior at the
Earth does not translate back into a luminosity distribution that varies with
distance.

\begin{figure*} \begin{center}
        \includegraphics[width=2.3in]{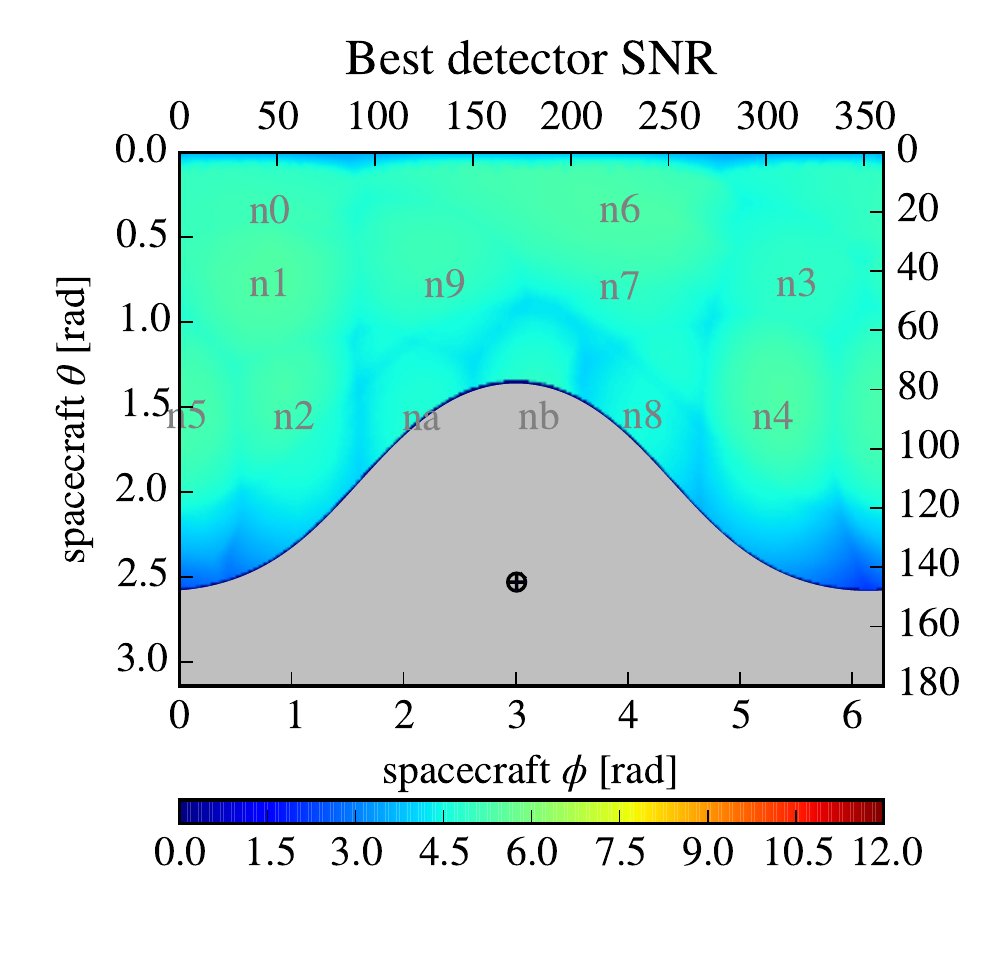}
        \includegraphics[width=2.3in]{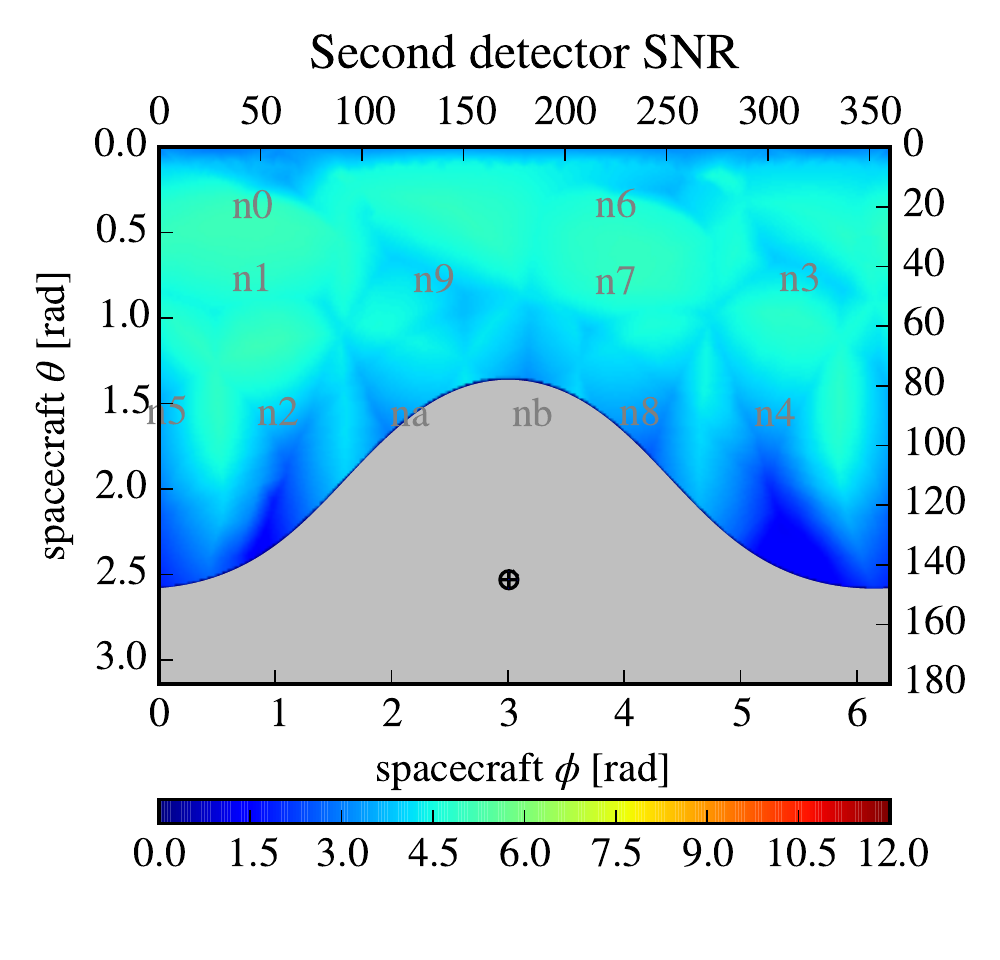}
        \includegraphics[width=2.3in]{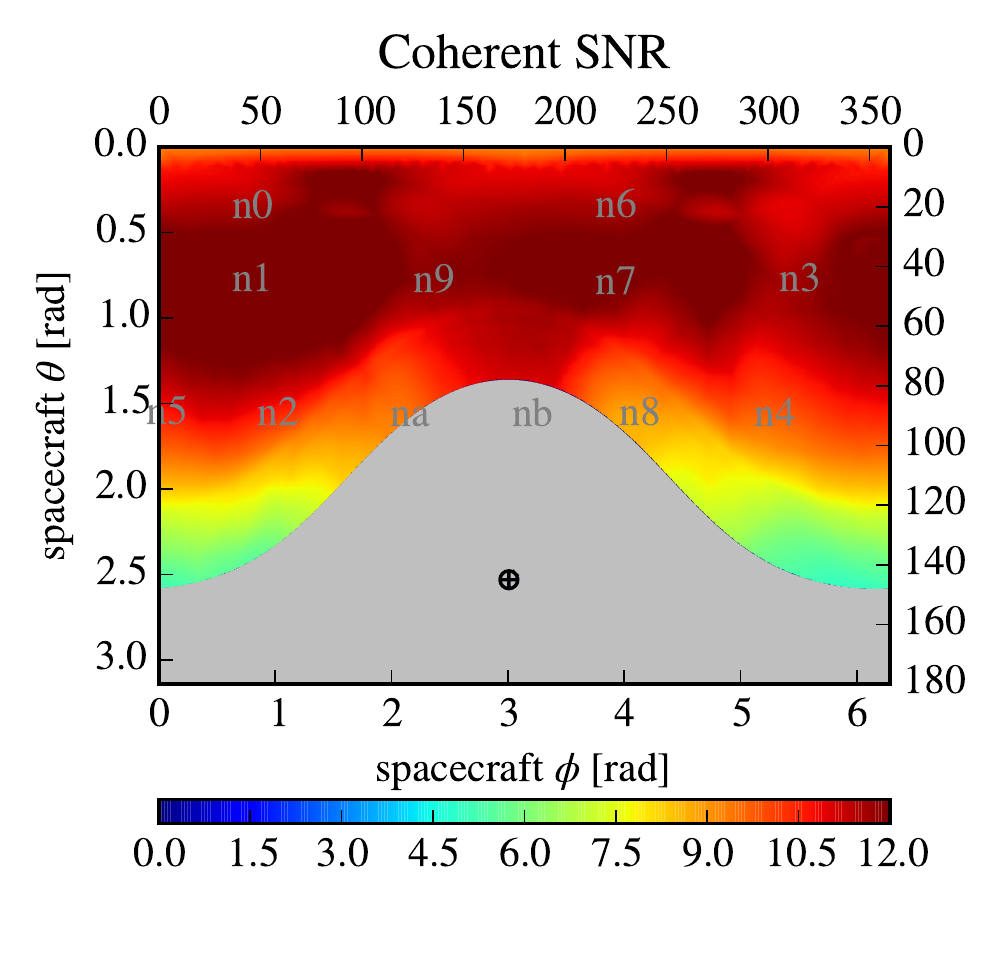}
        \caption{Signal-to-noise in GBM data expected from a 0.512s long signal with a
            normal GRB spectrum. The signal is normalized to 1.0
            photons/s/cm$^2$ in the 50--300 keV band, and the background rates
            and Earth position are selected from 10 seconds prior to GRB
            090305A. The model response includes contributions from atmospheric
            scattering. The first two maps show the signal-to-noise expected
            from the best and second-best detectors assuming data collected
            across 50--300 keV. The interpretation is as follows: given a
            hypothetical source at any sky location ($\phi$, $\theta$), what is
            the maximum (left) or second-best (middle) signal-to-noise seen
            across {\em all} NaI detectors.  Generally one would assume that the
            maximum SNR would come from the best-aligned NaI detector (n0, n1,
            \dots as shown), and the second-best SNR is from one that is
            nearby. The final plot shows the SNR expected from a coherent
            analysis of the data using all detectors, also as a function of
            source location. The maximum coherent SNR is achieved where
            multiple detectors are favorably oriented toward the source.  The
            coherent analysis samples all sky positions, which means it is
            subject to a trials factor, but this is difficult to evaluate
            because of the large degree of correlation between nearby sky
            locations. The ``best-detector'' SNR similarly has a trials factor
            equal to the number of detectors (12). Use of the ``second-best''
            detector, which corresponds to the on-board GBM triggering
            strategy, implies two detectors above a given SNR threshold, which
            means a much lower false-alarm probability at equivalent SNR, as
            well as improved rejection of non-Gaussian noise.}
        \label{fig:cohresp} \end{center}
\end{figure*}

One difficulty with any power-law prior is that it diverges for $s \to 0$.  We
enforce a finite and well-behaved prior by multiplying by a prefactor,
\begin{equation} P(s) = \left[1-e^{-\left(s/\gamma\sigl\right)^\beta}\right]
    s^{-\beta} \end{equation} so that $P(s)$ reaches a maximum constant value
of $(1/\gamma\sigl)^\beta$ for small $s$. The tunable parameter $\gamma$ sets
the number of standard deviations at which the prior begins to plateau, and we
use $\gamma = 2.5$.  We then approximate $P(s)$ as constant over a range of
$\sigl$ for any $s>0$, and include a correction to account for clipping of the
Gaussian for non-physical $s<0$, which can be represented by the error
function. The final approximation for the amplitude-marginalized log-likelihood becomes,
\begin{multline} \mathcal{L}(d) = \ln\sigl +
    \ln\left[1+\mathrm{Erf}\left(\frac{s_\mathrm{best}}{\sqrt{2}\sigl}\right)\right]
    + \mathcal{L}(d|s_\mathrm{best}) \\ + \left\{ \begin{array}{ll}
            \ln\left[1-e^{-\left(s_\mathrm{best}/\gamma\sigl\right)^\beta}\right]
            - \beta\ln{s_\mathrm{best}} & s_\mathrm{best} > 0 \\
            -\beta\ln\left(\gamma\sigl\right) & s_\mathrm{best} \leq 0
        \end{array}\right.  \end{multline} which contains factors from the
    Gaussian width, fractional overlap with $s > 0$, maximum likelihood at
    $s_\mathrm{best}$, and scaling from $P(s)$ respectively. Finally we are
    free to calibrate the log-likelihood by subtracting the expected
    $\mathcal{L}(d)$ calculated for no signal at a reference sensitivity:
    $\mathcal{L}_\mathrm{ref} = -\beta\ln\gamma +
    (1-\beta)\ln\sigma_\mathrm{ref}$. $\sigl$ represents the source amplitude
    required for a $1\sigma$ excess in the combined data, and is around 0.05
    photons/s/cm$^2 \times (T/1$s$)^{-1/2}$ [50--300 keV] for a typical source
    spectrum and background level. Figure \ref{fig:cohresp} shows the coherent
    signal-to-noise expected from all detectors for a 0.512s-long event with
    normal GRB spectrum and constant amplitude of 1.0 photons/s/cm$^2$, and
    compares it to the SNR expected from individual detectors alone in the
    50--300 keV band.

\subsection{GBM prompt coincidence with GW events}


A GW trigger provides an accurate time of coalescence $t_c$ as
well as an approximate sky location to tens or hundreds of square degrees. This
matches well the time and sky location constraints provided by a prompt
observation by GBM, so that coincidence between the two instruments is
particularly effective. To make use of the coherent analysis described in the
previous section, we must first obtain an expected all-sky GBM response for a
given counterpart. We make use of three representative source spectra using a
set of precomputed response tables. The representative spectra are Band
functions with the parameters listed in table~\ref{tab:bandspectra}.
\begin{table} \begin{center} \begin{tabular}{cccc} spectrum & $E_\mathrm{peak}$
            & $\alpha$ & $\beta$ \\ \midrule normal & 230 keV & 1 & 2.3 \\ hard
            & 1 MeV & 0 & 1.5 \\ soft & - & 2 & - \end{tabular} 
\caption{Model Band spectral parameters used to generate GBM detector response
tables used in the coherent analysis. The soft and hard spectra were originally
designed to produce a representative response over the primary GBM-GRB localization band 50--300
keV, and a number of more realistic response models are currently being
developed which will be more approriate for full-spectral all-sky analysis.
The construction of these all-sky, all-time models is difficult
computationally due to the many degrees of freedom in the detector response.}
\label{tab:bandspectra}
\end{center}
\end{table}
For the soft spectrum, $E_\mathrm{peak}$ and
$\beta$ are undefined. The tables provide channel-dependent expected counts for
the direct and spacecraft-scattered response as a function of source location
to $\sim$1$^\circ$ resolution. Contributions from atmospheric scattering are
available as counts summed over the two CTIME channels covering 50--300 keV as
a function of source and Earth relative position for the two most common
spacecraft rocking angles.

For each GW candidate event, we search a window \mbox{[--30s, 30s]}
relative to $t_c$ for prompt excess between 0.256s and 8s long. Emission
outside of the standard accretion timescale covering the first seconds
following merger is speculative. However, any events detectable in
gravitational waves will be much closer than known sGRB's to date, and
therefore they provide an opportunity to search below threshold for weak and
possibly less-beamed precursor or extended emission. To appropriately tile the
search in time and duration $T$, we use rectangular windows with $T$ spaced by
powers of two (0.256s, 0.512s, 1s, etc.).  Their central times are sampled
along the search interval in units of $T/4$ to provide an even mismatch in
signal-to-noise across search windows. For each emission model tested, the
likelihood ratio is then marginalized over all windows and spectra.

The GW data also provides a rough sky-location, which can be
represented as a prior probability distribution over the sky
$P_\mathrm{GW}(\Omega)$. This GW prior is multiplied by the GBM
likelihood ratio, which is also a function of sky position, before
marginalization over sky location is done,
\begin{equation} \label{eqn:lgwem} \Lambda_{\mathrm{GW-GBM}} = \int
    d\Omega\,P_\mathrm{GW}(\Omega)\,\Lambda_\mathrm{GBM}(\Omega) \end{equation}
where the GBM likelihood ratio depends on source spectrum, on-source time
window, and location. While the sky prior can in principle be improved by
incorporating knowledge of the anisotropic local mass distribution (using a
galaxy catalog), we note that as neither the GW network nor GBM have precise
(sub-degree) sky localization, the gains from sharpening the prior in this way
are small, and catalog incompleteness and luminosity-rate relationships
introduce unnecessary complications. The situation is different when
considering better-localizated X-ray (and optical/ratio) counterparts.

\subsection{GW-GBM coincident background estimation}

The background for GW-GBM analysis is characterized by the probability of
observing a given $\Lambda_\mathrm{GW-EM}$ from random coincidence between GW
and GBM data.  We calculate the expected distribution of
$\Lambda_\mathrm{GW-EM}$ by running the coincidence search defined by equation
\ref{eqn:lgwem} on GW background events derived from time-shift analysis. The
time-shift sample represents 100 artificial time-shifts ($\sim$sec) applied to
data between the LIGO Hanford and Livingston sites, as well as a further 10
time-shifts ($\sim$day) applied between the GW and GBM data. Thus we obtain a
background sample representing characteristic noise from $\times 1000$ the
original foreground live-time.

\Needspace*{2\baselineskip}
\subsection{Simulation of prompt gamma-ray counterparts}

GRB simulated signals are injected into the data by using the instrument
response model to predict the signal contribution from a source with a given
spectrum and fluence. In our analysis, we simulate a standard candle signal
following the normal-type Band spectrum with a fluence at the Earth of 1
photon/cm$^2$/s in 50--300 keV assuming a source at 30 Mpc. The signal assumes
normal GRB spectral parameters according to table \ref{tab:bandspectra} and
lasts for one second, corresponding to an isotropically emitted energy of about
$6.6\times 10^{46}$ erg. When simulating a signal at distance $D$, the fluence
is reduced by a factor of $(30\text{ Mpc}/D)^2$.

\Needspace*{4\baselineskip}
\section{X-ray counterparts with RXTE ASM}

\subsection{ASM instrument and data products}


The All-Sky Monitor (ASM) \citep{Levine:1996du} on the Rossi X-ray Timing
Explorer (RXTE) surveyed the X-ray sky between 1.5--12 keV from 1996--2011. The
instrument consisted of three long-and-narrow shadow-mask X-ray cameras rigidly
attached to a common motorized axis of rotation. Each camera saw a 6$^\circ$ by
90$^\circ$ field-of-view, and together covered about 3\% of the sky at any one
time. As the cameras scanned across the sky while tiling 90-second dwells, they
were able to localize a steady source to about 0.1$^\circ$. The duty cycle of
the ASM along with constraints from the observing schedule of other RXTE
instruments resulted in a randomly chosen location being scanned by an ASM
camera a few times per day.

An ASM camera observation consists of a superposition of the shadow pattern of
all X-ray sources within the field-of-view. The data can be modeled as the
contributions from active sources with known locations along with a diffuse
background. Amplitudes for sources are treated as free parameters, and
estimated using a linear least-squares fit to the data \citep{Levine:1996du},
with errors dominated by photon counting statistics. The amplitude from any
additional position of interest within the field-of-view can also be estimated
by adding the location as a source during the fit. For a weak source, the error
on estimated amplitude is dominated by the diffuse background which typically
contributes a 3$\sigma$ error of $\sim$20 mCrab ($4.8\times10^{-10}$ erg/cm$^2$/s
between 2--10 keV).

\subsection{Modeled search for afterglow light-curves}


The ASM measurements consist of irregularly-spaced flux observations with
varying uncertainties from specific regions in the sky, which makes a search
strategy for an arbitrary flux excess complicated. We use limited knowledge of
the sGRB X-ray afterglow signal to narrow the search to signals that decay like
broken power-laws in flux beginning from the initial time-of-merger $t_0$ as
determined by the GW data.

\Needspace*{2\baselineskip}
Following \cite{Zhang:2005fa}, we model a canonical sGRB X-ray afterglow
by a double-broken power-law with a short region of rapid decline (power law
index $\alpha \approx 3$) lasting 1e2--1e3 seconds, followed by a standard
afterglow decay ($\alpha \approx -1.2$), and finally a break and raid decay
($\alpha \approx -2$) after 10$^4$--10$^5$ seconds. Additionally there may be a
plateau or extended emission somewhere after the rapid decay phase around
10$^2$--10$^3$ seconds. We attempt to create a generic light-curve with minimal free
parameters which covers possible observed light-curve scenarios, in context of
the limited sampling of ASM data, and also allows for the possibility of
delayed or non-standard X-ray emission due to off-axis observation or some
other unobserved phenomenon.

\begin{figure}
\begin{center}
    \includegraphics[width=3.4in]{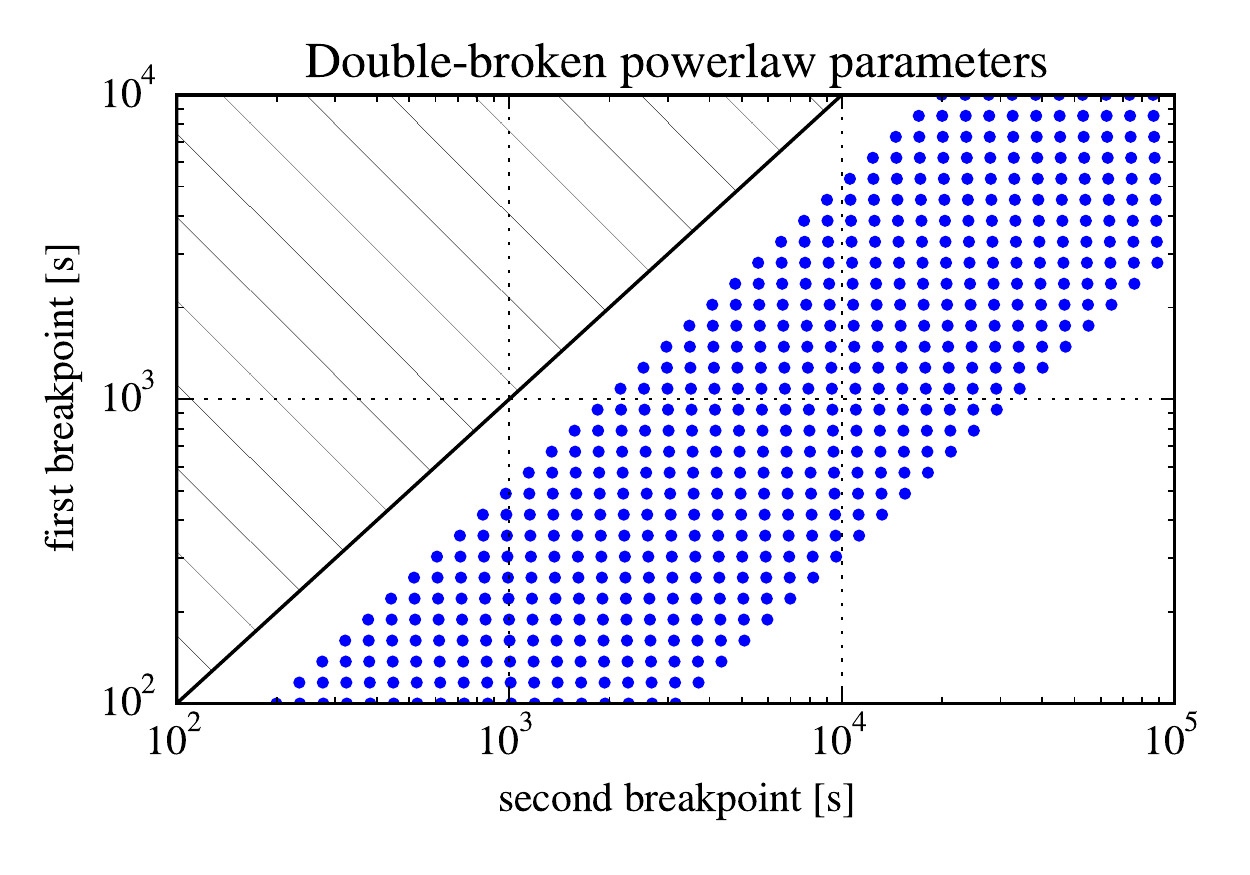}
    \includegraphics[width=3.4in]{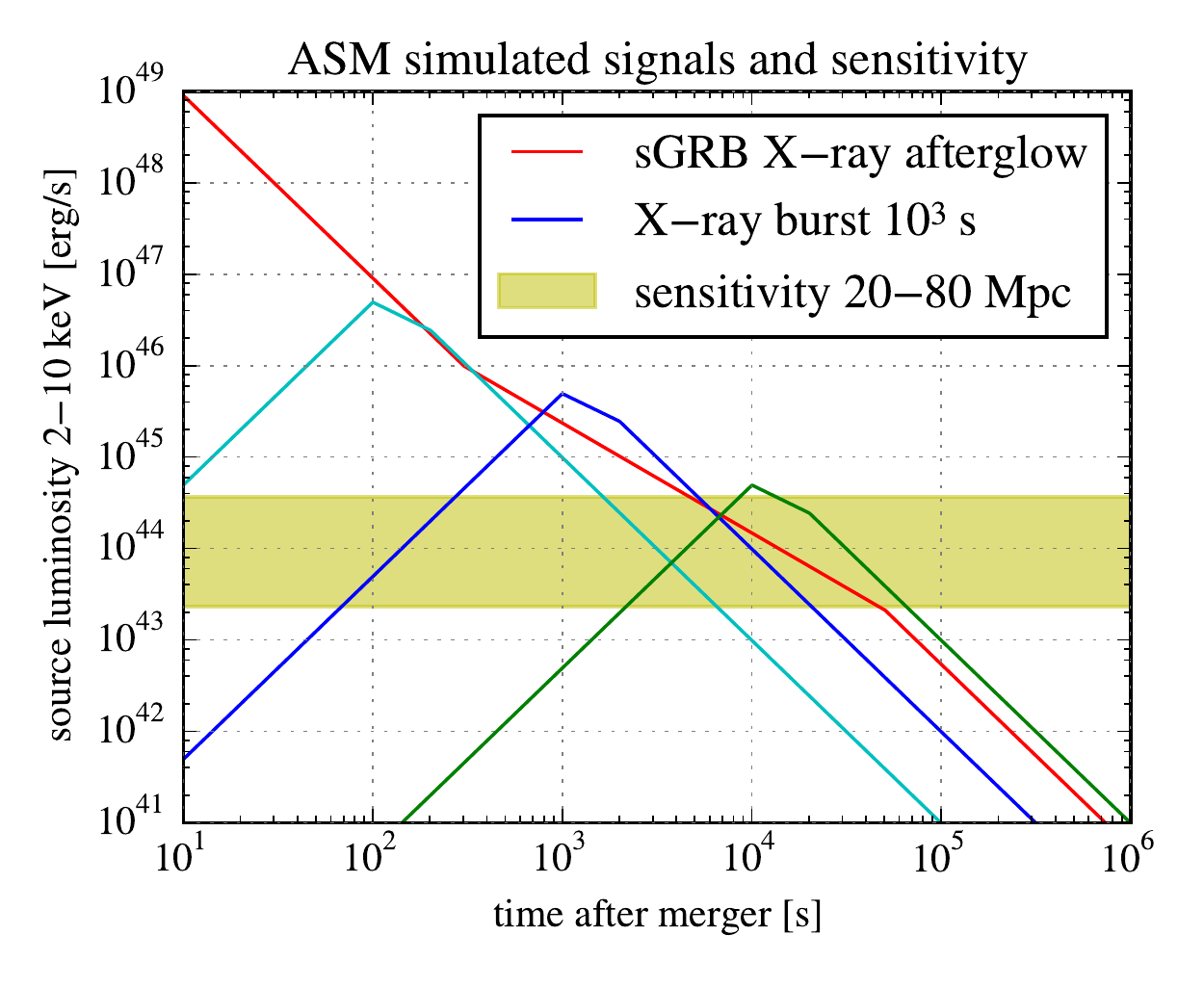}
    \caption{(top) double-broken
power-law break-point parameters for model afterglow light-curves used when
fitting ASM data. (bottom) typical 2--10 keV sGRB afterglow light-curve with
decay indexes of $-$2, $-$1.2, $-$2 between breakpoints at $3\times10^2$ and
$5\times10^4$ seconds. The amplitude is based on Swift XRT observations of
actual sGRB afterglows, which can vary in intrinsic luminosity by over two
orders of magnitude. Also shown is a family of ad hoc \mbox{X-ray} bursts, which have
a total 2--10 keV energy release of $10^{49}$ erg over their durations. This
amplitude is chosen so that the bursts at peak amplitude rise just above the
afterglow signal, as in the case for the extended emission seen in some sGRB
afterglows. The bursts are also parameterized by a double-broken power-law,
this time with decay indexes of $+$2, $-$1, $-$2 between breakpoints at $T$ and $2T$
where $T$ is the duration. The yellow band represents an intrinsic X-ray
luminosity required to be detectable ($>$20 mCrab) in a single 90s dwell by ASM
for sources between 20 and 80 Mpc.}
\label{fig:afterglowlc}
\end{center}
\end{figure}

For this we sample double-broken model power-law light-curves with
break times sampled on a grid (figure \ref{fig:afterglowlc}) and freely-varying
power law indices. We label $N$ local measurements of flux excess $d_i$ by ASM
at times $t_i$ with standard deviations $\sigma_i$. A given light-curve
parameterization defines expected relative counts at the measurement times $L_i
= L(t_i)$. We can then define the weighted sum of measurements $c_i$ which
maximizes the signal-to-noise of the summed data, \begin{equation} D = \sum_i
    \frac{L_i d_i}{\sigma_i^2}.  \end{equation} The measurements $d_i$ are
already assumed to be zero-mean, giving a measured signal-to-noise for the
coherent sum $D$ of, \begin{equation} \rho = \frac{D}{\sqrt{\sum_i L_i^2 /
            \sigma_i^2}} \end{equation} The expected signal-to-noise of each
measurement is added in quadrature for the total expected signal-to-noise of
the coherent sum, \begin{equation} \E{\rho} =
    \sqrt{\sum_i\frac{L_i^2}{\sigma_i^2}} \end{equation} If we are fortunate to
have an observation soon after the beginning of the afterglow, the first
measurement is likely to dominate the sum due to the rapid decay. If only
late-time measurements are available, the coherent sum naturally bins the data
in order to gain in sensitivity. A positive power-law index for the first
segment of the light-curve can handle situations where the afterglow does not
begin immediately after the burst, for example due to beaming effects.

For each pair of power-law breakpoints ($t_1$, $t_2$), we fit the three
power-law indices ($\alpha_1$, $\alpha_2$, $\alpha_3$) using a minimization
routine under the constraints $-3 < \alpha_1 < 2$, $-2 < \alpha_2 < 1$, and $-3
< \alpha_3 < 0$. This allows for a wide variety of afterglow waveforms
including the standard canonical double-broken power-law, an initially rising
light-curve which then decays normally, or an isolated burst of extended
emission that might occur minutes to days after merger. For each set of ASM
measurements from a location of interest [0, 4d] about $t_0$, we record the fit
parameters which give the maximum SNR, as well as a variety of auxiliary
parameters useful for characterizing non-Gaussian noise.

\subsection{ASM afterglow coincidence with GW events}

The ASM point source flux estimation is well-suited to follow up locations of
likely host galaxies for GW events. The shadow-mask flux
reconstruction isolates contributions to about 0.1$^\circ$ on the sky. This is
small enough that a large fraction of the sky can be excluded by requiring
known galaxy coincidence, but also large enough so that all but the most
extended nearby galaxies can be targeted by a single location. For each host
location, we generate ASM flux measurements from available camera dwells by
adding that single location as a test source along with other known active
X-ray sources during the fit of each dwell's shadow-mask data. Derived flux
measurements from that location are then searched numerically for the five
parameter light-curve (two breakpoints and three decay factors) which maximize
SNR of the coherent sum. Because the data are sparse, it is difficult to
maximize SNR numerically for freely-varying breakpoints, so they are sampled on
a fixed grid and only the three power-law indices are searched with standard
numerical minimization routines as described above.

\begin{figure} \begin{center} \includegraphics[width=3.4in]{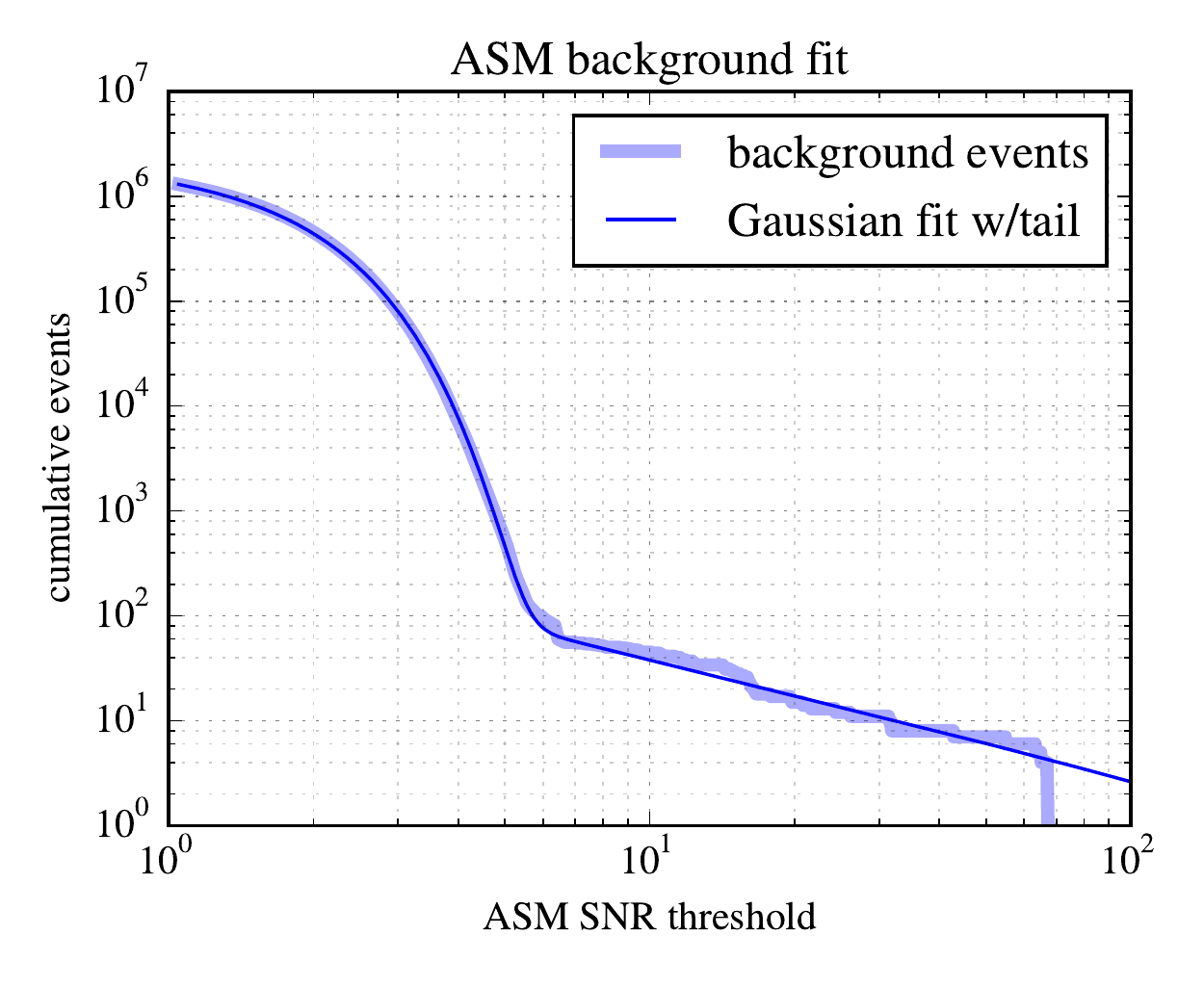}
        \caption{All-sky X-ray background from the generic light-curve fitting
            procedure. The background distribution is largely Gaussian, but
            shows a significant tail which can be captured by a power-law. The
            background fit is used to estimate the significance of coincident
            signal-to-noise measurements $\rho$ in the analysis.}
        \label{fig:asmbg} \end{center} \end{figure}

We take the maximum signal-to-noise $\rho$ observed across all model
light-curves to weigh each candidate host galaxy according to likelihood. The
conversion from $\rho$ to likelihood is determined by a global SNR distribution
of X-ray background derived from the same methods. The distribution is fit
to a function $f(\rho)$ parameterized by a Gaussian (maximum value 1) with
power-law tail (figure~\ref{fig:asmbg}), and this empirical fit is used to
calculate the likelihood = $1/f(\rho)$ at each SNR measurement. For a given
GW event with a set of possible hosts, we obtain a final rank
for the coincident observation,
\begin{equation}
    \Lambda_\mathrm{GW-ASM} =
    \sum_\mathrm{host}\left[ P_\mathrm{GW}(\mathrm{host}) \times
        f(\rho_\mathrm{host})^{-1}\right]
\end{equation}
where the likelihood $\Lambda_\mathrm{GW-ASM}$ represents the probability of an
afterglow signal in the ASM data, marginalized over all possible hosts. The
hosts must maintain a $P_\mathrm{GW} > 1$ (a threshold representing the case
for a galaxy of typical $\sim$MW mass and no location or distance prior
information), and at maximum 200 hosts are scanned to limit computational
expense.  $P_\mathrm{GW}(\mathrm{host})$ is derived, as in equation
\ref{eqn:gwgcprobability}, from the sky and distance overlap of each host with
the prior distributions determined by the GW data and host mass
(luminosity). To convert the likelihood into a more SNR-like quantity, we
define rank $r = \sqrt{2 \Lambda_\mathrm{GW-ASM}}$.

\subsection{Simulation of X-ray afterglow counterparts}

\begin{figure} \begin{center}
        \includegraphics[width=3.4in]{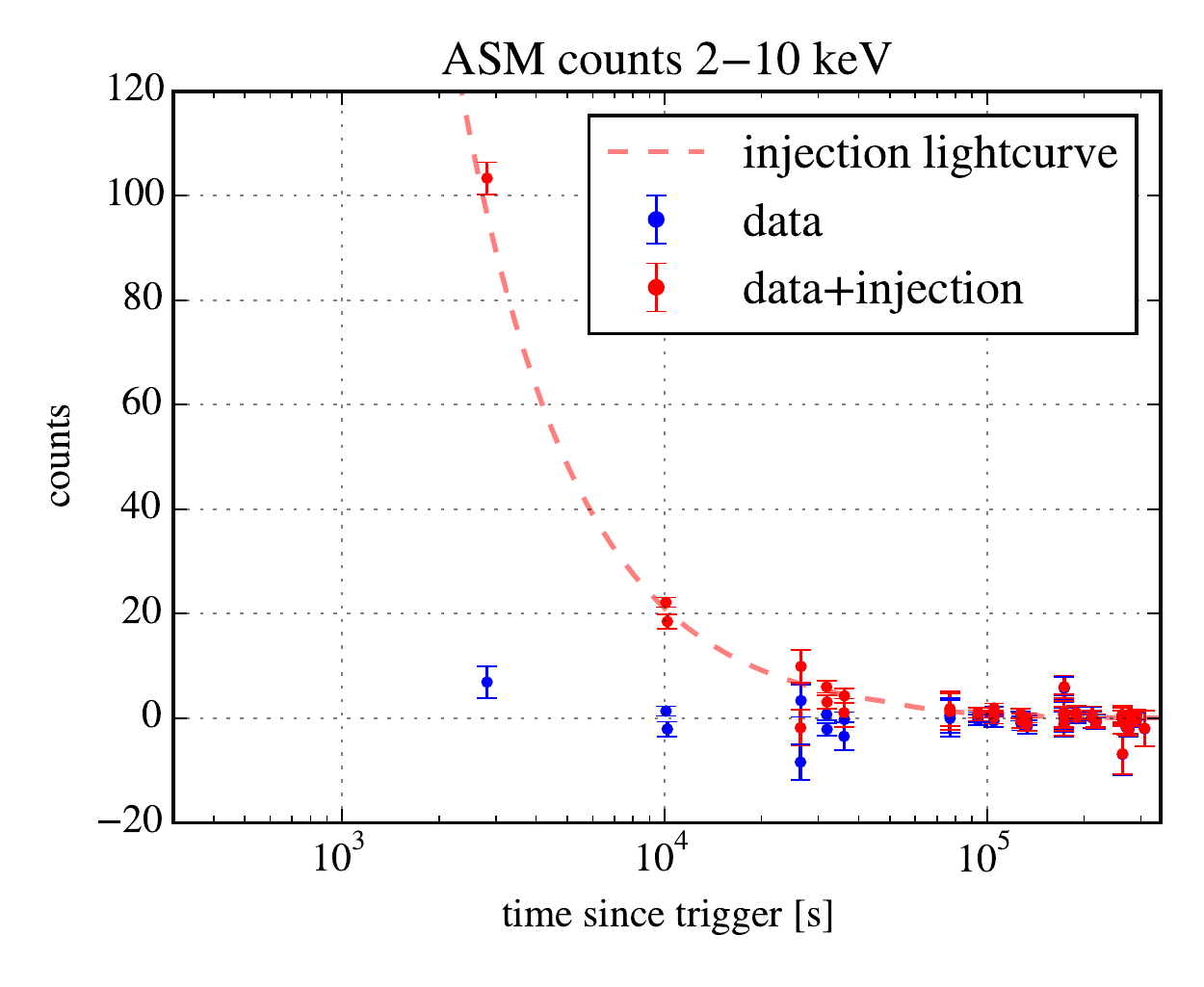} \caption{ASM data
            with injected afterglow light-curve at 14 Mpc. The best fit
            light-curve provides the optimal coherent sum signal-to-noise
            measurement.}
\label{fig:asminjection}
\end{center} \end{figure}

We simulate a family of X-ray afterglow signals by adding model light-curves
directly to the derived ASM flux measurements for a given location. We assume
that the error estimate for each flux measurement remains dominated by
contributions from the diffuse background, so that Poisson or other systematic
error from the source can be neglected. The model light-curve is then injected
into the data from a reconstructed location at a time of interest (figure \ref{fig:asminjection}). For a given
light-curve shape, we recover the distribution of maximum SNR measurements from
the template search as a function of injected amplitude, or equivalently
injected distance for a standard candle source.

For the case of a GW-ASM coincidence, we begin with a set of simulated NS/NS
coalescence events detected with the GW analysis
procedure. For S6/VSR2+3, a large number of simulations were done at random sky
locations and distances in order to evaluate the overall detection efficiency
of the pipeline \citep{Colaboration:2011nz}. To characterize the sensitivity of
the joint GW-ASM coincidence analysis which assumes signals originating from a
set of host galaxies, we first create an artificial galaxy at the location and
distance of the GW simulation, with a luminosity taken from a
(luminosity-weighted) random draw of galaxies from the catalog within some
compatible volume. We then synthesize ASM data from this fake location and
distance by appropriating the ASM flux measurements from the randomly chosen
galaxy. Reconstructed ASM flux measurements from the remaining true galaxies
within the GW-derived sky region are included in the coincidence analysis as
before, as well as the fake galaxy with the simulated lightcurve added to the
ASM data. We then obtain a distribution of the joint GW-ASM detection statistic
for a given CBC system as a function of distance and intrinsic EM amplitude.

We test performance of the method against a standard X-ray afterglow
typical of Swift XRT observations of those from short GRBs (figure
\ref{fig:afterglowlc}). This light-curve has power-law decay indices of $-2$,
$-1.2$, and $-2$ between breakpoints at $3 \times 10^2$ and $5 \times 10^4$
seconds, and a 1--10 keV luminosity of 10$^{46}$ erg/s at the first
break-point.  We also assume a Crab-like spectrum, for convenient conversion
into ASM band counts which are Crab calibrated.

In addition, we inject a representative sample of X-ray burst signals which
occur minutes to days after the merger, motivated by theoretical magnetar
wind-driven scenarios \citep{Zhang:2012wt, Gao:2013rd}. We test a short,
medium, and late-time burst beginning with a rise $\propto t^2$ until $t_1$ =
10$^2$, 10$^3$, and 10$^4$ seconds, decaying as $t^{-1}$ until $t_2 = 2t_1$,
and then falling as $t^{-2}$ afterward. The simple wind-driven emission
scenarios consider a total amount of energy released as the magnetar spins
down, and then either dissipating via internal or external shocks, releasing
some fraction of the dissipated energy in X-rays.
We set the total energy released in X-rays over the entire period of the burst
to be $10^{49}$ ergs in 2--10 keV, resulting in bursts with instantaneous
luminosities slightly greater than that for the standard afterglow at peak. In
this way, they can be thought of representing the extended emission/flares seen
in some sGRB afterglows. As in the standard afterglow model, we assume a
Crab-like spectrum for convenient translation into ASM counts.


\subsection{GW-ASM coincident background estimation}

The GW-ASM background distribution is found by running the coincidence search
using GW background events derived from time-shift analysis, similar to that
for GBM. The same 10 additional time-shifts (spaced in multiples of 4 days) are
applied between the GW and ASM data to increase statistics. We then obtain the
distribution of the GW-ASM rank statistic expected from accidental coincidence
between GW and ASM background. This distribution, as well as that for the family
of simulated afterglow signals, are shown in figure \ref{fig:asmrank}.
 

\Needspace*{4\baselineskip}
\section{Results and conclusions}

\subsection{Selection cuts and background rejection}
\label{sec:selectioncuts}

\begin{figure} \begin{center}
        \includegraphics[width=3.4in]{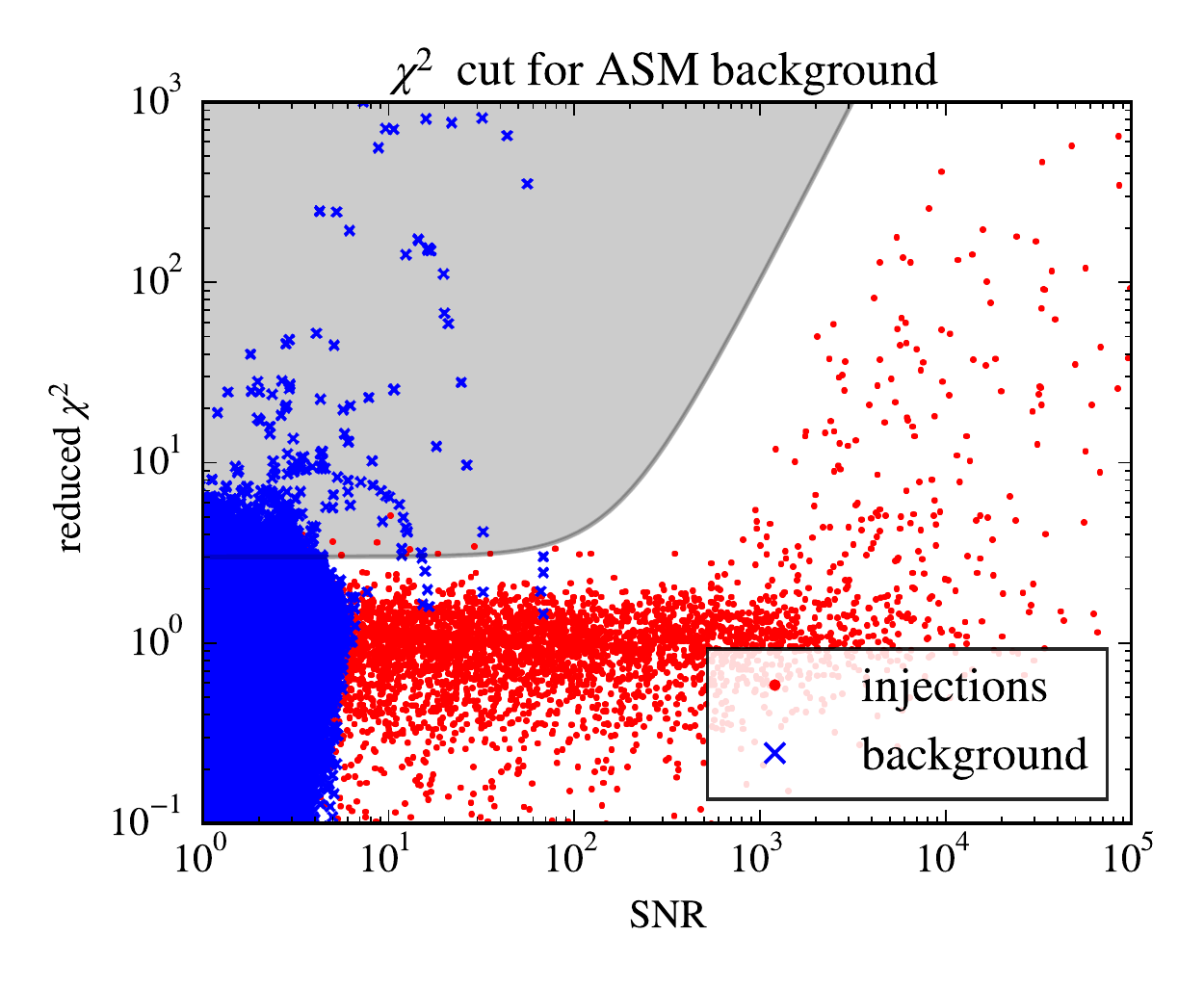}
        \caption{Reduced $\chi^2$ vs max-SNR in ASM data of background and
            simulations (all waveforms) for GW-ASM events. The scatter points
            represent full end-to-end implementation of the coincident GW-ASM
            pipeline, including joint simulation of NS/NS and afterglow signals
            from nearby host galaxies. We cut at $\chi^2 > 3$ (gray shade) to
            remove loud background events inconsistent with one of the template
            waveforms.  For example, they may have excess flux before the
            merger time, or sporadic flux afterward inconsistent with a
            power-law decay. The cut is relaxed at high SNR due to template
            mismatch.} \label{fig:asm1} \end{center} \end{figure}

We have tested the end-to-end GW-EM pipelines for both GBM and ASM on a
selection of real GW background noise events from the LIGO-Virgo S6/VSR3 joint
science run, as well as simulated GW-EM events injected in coincidence in both
the GW and EM data streams. By comparing the properties of GW background and
simulations, we identify a number of selection cuts that help reduce the
non-Gaussian background present in the EM data. The ASM selection cut is based
on the reduced $\chi^2$ of the ASM data after the best-fit double-broken
power-law light-curve has been subtracted (figure \ref{fig:asm1}). However, due
to the sporadic coverage of the ASM measurements, it is difficult to select on
any morphological properties of the light-curves themselves, beyond the weak
requirement that the measurements be consistent with one or more of the
possible templates.

\begin{figure}
    \begin{center}
        \includegraphics[width=3.4in]{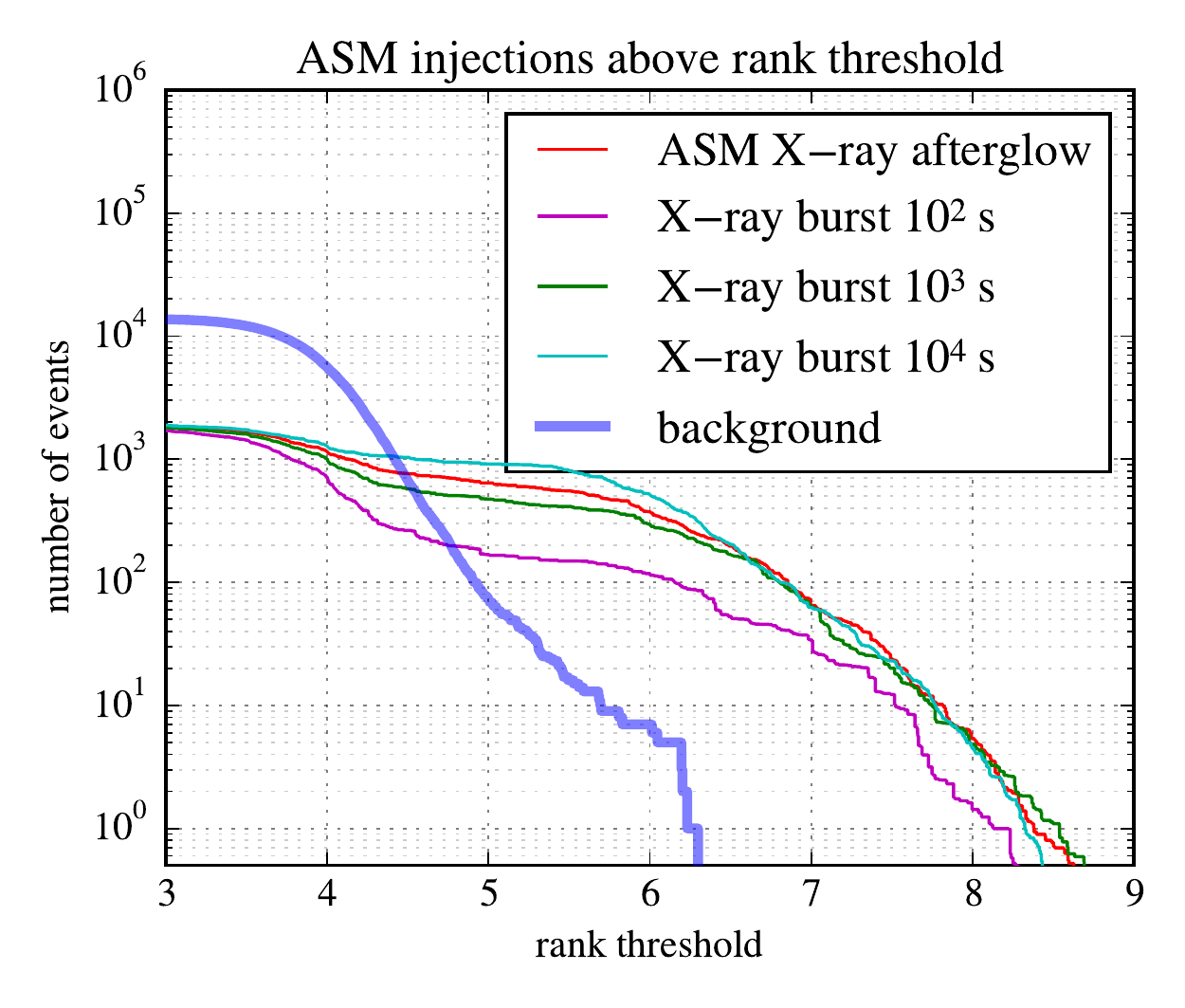} \caption{Cumulative
            distribution of ASM rank (monotonic with GW-ASM likelihood ratio)
            for background events, as well as a variety of GW-ASM afterglow
            simulations, volume-weighted to resemble a spatially homoegeneous
            distribution in distance up to 40 Mpc. The standard afterglow
            follows the parameterization in figure \ref{fig:afterglowlc} with
            an intrinsic luminosity of $10^{46}$ erg/s at the first breakpoint
            at 300s. The other three simulated afterglows represent X-ray
            bursts at various timescales, with characteristic onset and
            duration of 100, 1000, and 10000s, and a total integrated energy
            release of $10^{49}$ erg.}
    \label{fig:asmrank}
    \end{center}
\end{figure}

\begin{figure} \begin{center}
        \includegraphics[width=3.4in]{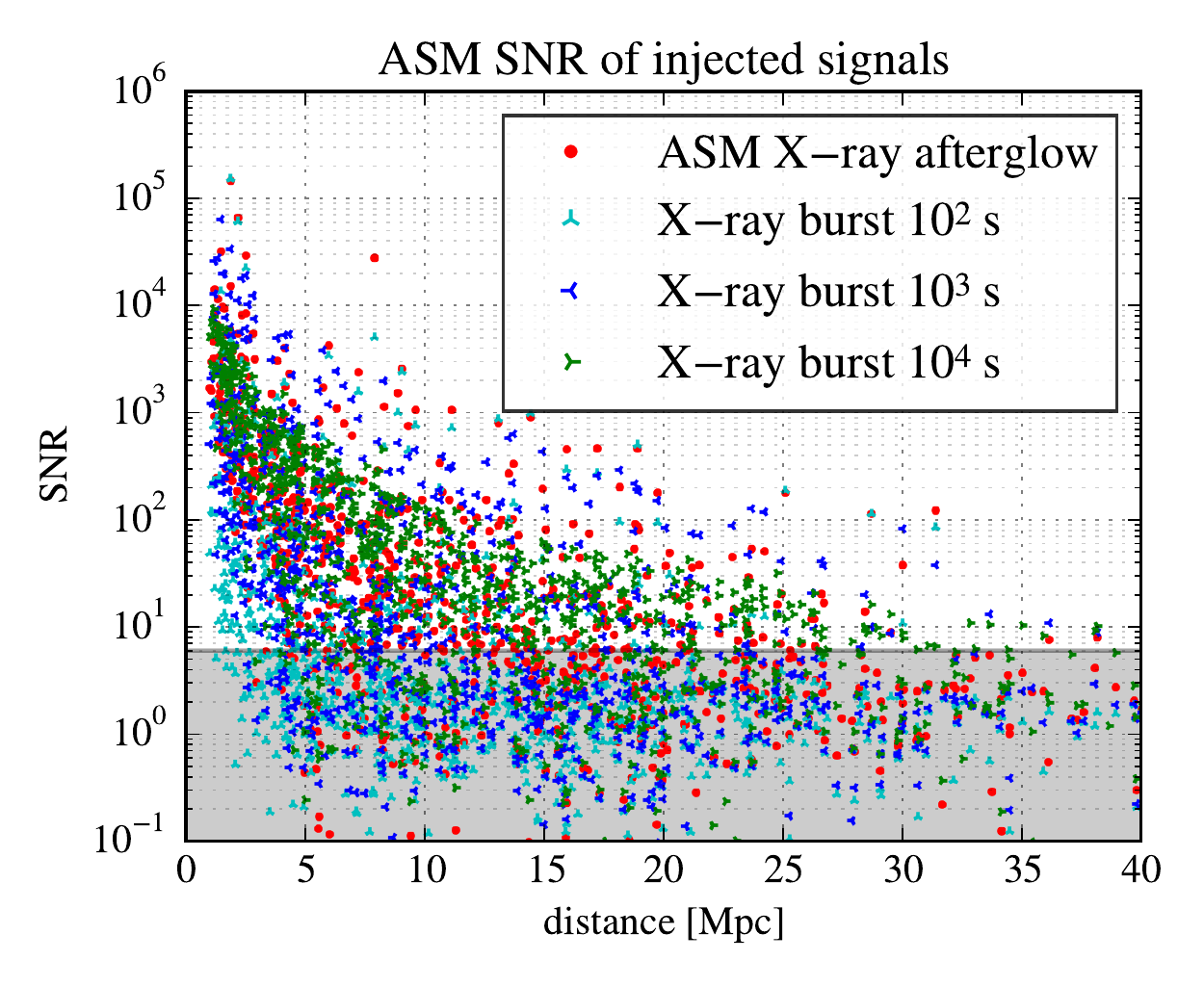} \caption{Best-fit SNR of
ASM injections versus distance for joint GW-ASM NS/NS-afterglow injections
split by model waveform. The ASM SNR is highly variable mainly due to the
uncertain time-to-first observation. Missed detections in ASM are due to
several factors including lack of ASM data or poor sky localization due to
antenna pattern effects. Rather than cut directly on SNR, ASM events are
selected using the derivate {\em rank} parameter (figure \ref{fig:asmrank})
which folds in SNR information from all possible host locations and their
probabilities, as well as information about the ASM background distribution.
For reference, we show a shaded region at $\mathrm{SNR}<6$ which encompases the
majority of ASM background (figure \ref{fig:asm1}).} \label{fig:asm2}
    \end{center} \end{figure}

\begin{figure} \begin{center} \includegraphics[width=3.4in]{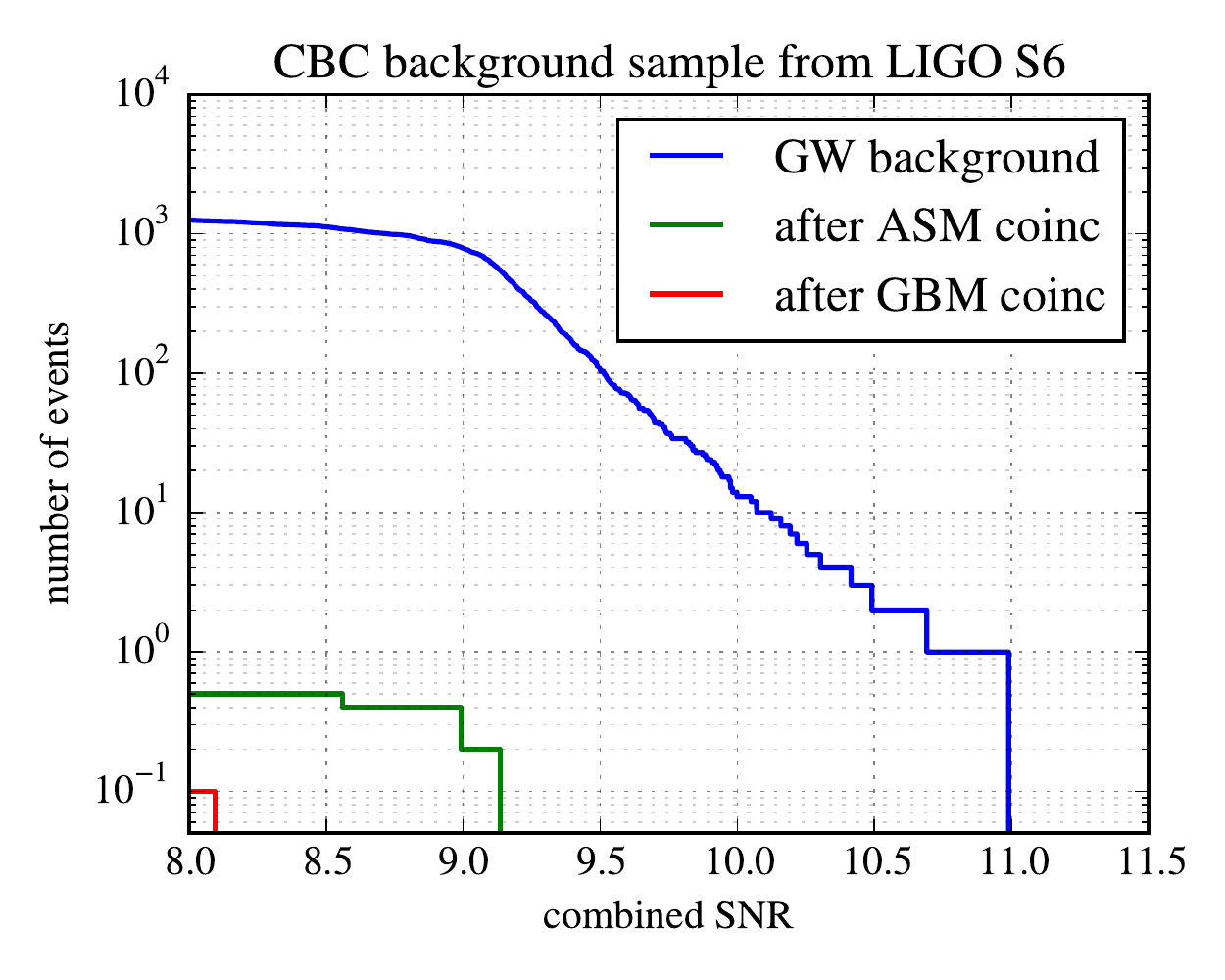}
        \caption{Cumulative distribution of combined SNR ($\rho_c$)
            for time-shifted background events observed in GW data before and after selection cuts
            are made on the requirement of an EM coincidence from either GBM or
            ASM.
An additional 10 time-shifts are applied between GW and EM data, resolving
the expected distribution after coincidence to 0.1 events. The coincidence rejection factors are $\sim$$10^{-3}$ and
            $10^{-4}$ for ASM and GBM respectively. The corresponding loudest
            events are at a combined SNR of 11.0, 9.1, and 8.1. However below a
            combined SNR of $\sim$9, the effects of other analysis cuts take
            effect. Additional factors of background rejection from tighter
            time and sky coincidence could further dig into the noise
            distribution as suggested in \cite{Camp:2013cwa}, but demonstrating
            that robustely would require a much larger, or lower-threshold GW
            background set than was used this study.}
        \label{fig:cumhist}
    \end{center} \end{figure}

We place a nominal threshold of 18 on $\Lambda_\mathrm{GW-ASM}$, the GW-ASM
coincident likelihood ratio. Because the likelihood is largely determined by
the Gaussian-like shape of the ASM background distribution, we represent this
as a threshold of 6 on the alternative rank parameter $r=\sqrt{2\Lambda}$.  The
distribution of $r$ for the GW background events, as well as a variety of
simulations is shown in figure \ref{fig:asmrank}. With the ASM coincidence
requirement applied, the loudest GW background event drops from a combined SNR
of 11.0 to 9.13 (figure \ref{fig:cumhist}), a reduction of $\sim$17\%. ASM is
generally sensitive to afterglow signals within the LIGO horizon, however
detectability varies widely since it is highly dependent on exactly when, or if
ever, the ASM observation(s) occur relative to the peak in the light curves
(figure \ref{fig:asm2}). Similarly, detections may fail if the GW localization
is not sufficient to pick out the correct galaxy host among the top 200
candidates, which happens about 7\% of the time at 10 Mpc, and 32\% of the time
at 30 Mpc.

In the GBM detector, most short background events are due to phosphorescent
events in the NaI detectors, which can arise from cosmic rays. While we
implemented a rough cut to exclude such events from contaminating the
background fits during the analysis, we implement a further selection to remove
remaining particle events from the foreground interval. The particle events are
unique in that they show up in a single detector, and are generally soft in
reconstructed energy. Thus we implement two cuts based on the ratio of detected
channel 0 SNR in the loudest and second-loudest NaI detectors, as well as
between channel 0 and channel 1 in the loudest detector. With these cuts we are
able to remove most of the remaining soft particle background.

\begin{figure} \begin{center}
        \includegraphics[width=3.4in]{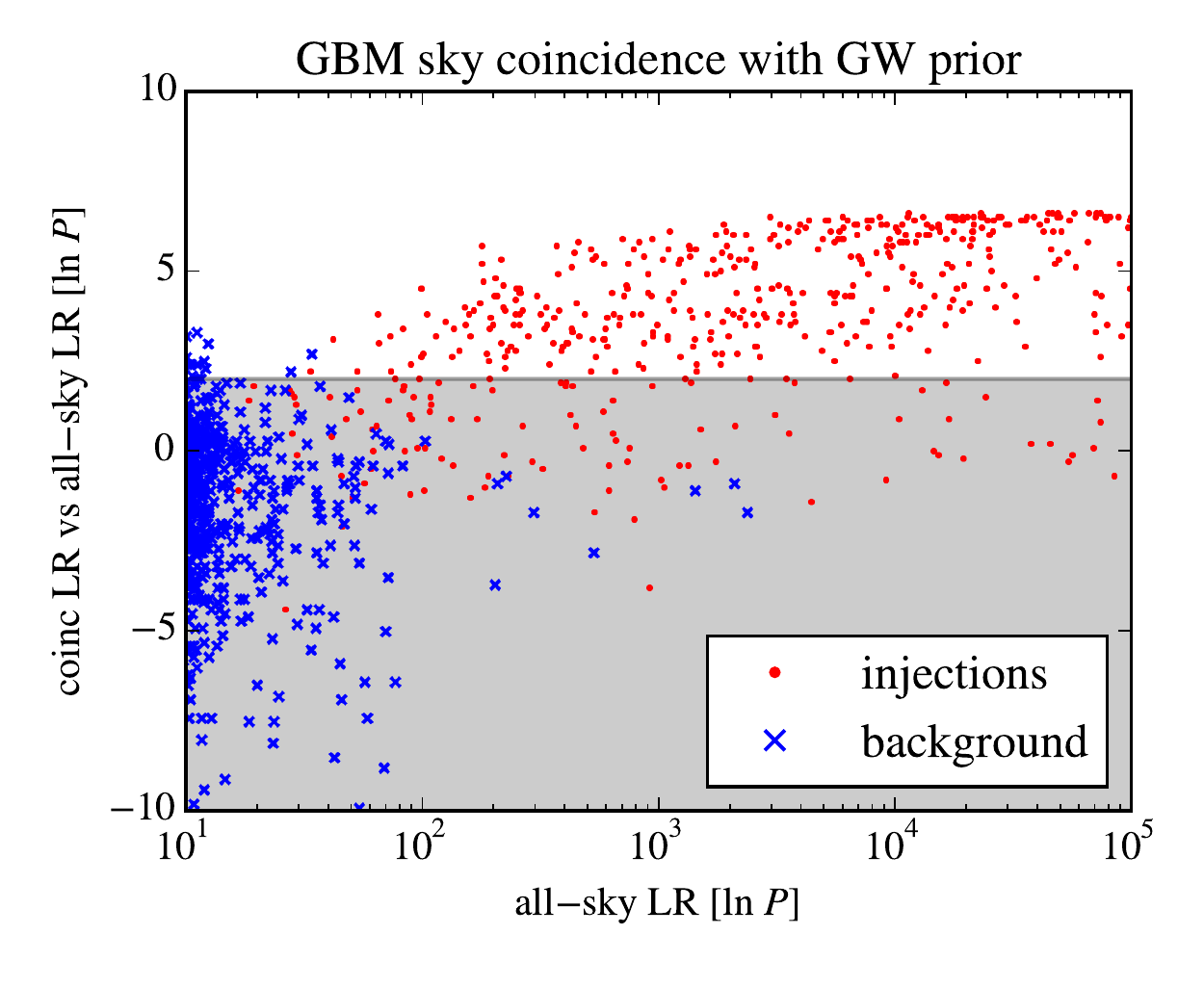} \caption{Sky
            coincidence cut for GW-EM search. We require the likelihood ratio
            of GBM signal vs no-signal to be a factor of $e^2$ greater when
            the GW skymap prior is used, versus an all-sky prior. The
sky-coincidence is effective at removing loud GBM background,
although some signal injections
are also rejected due to poor GW or GBM sky localization.}
        \label{fig:gbm2} \end{center} \end{figure}

We also implement a sky-coincidence cut between the GW sky location and GBM by
comparing the GBM likelihood ratio with and without the GW sky prior folded in.
Since the sky prior typically covers a fractionally small region of the sky for
a well-localized event (100s of square degrees), we expect an appropriately
normalized coincident observation to have a correspondingly higher likelihood
ratio than one that assumes an isotropic all-sky prior. We expect a factor that can
be roughly in line with the fractional reduction in sky area ($\sim$400), but
since this can be moderated by many effects (averaging windows, systematic
errors in reconstructed location), we choose an empirical factor of $e^2$ as a
coincidence requirement. This cut is still able to reject a large fraction of
the background where we do not expect any systematic increase in likelihood
ratio due to incorporation of a GW sky prior (figure \ref{fig:gbm2}).

\begin{figure} \begin{center} \includegraphics[width=3.4in]{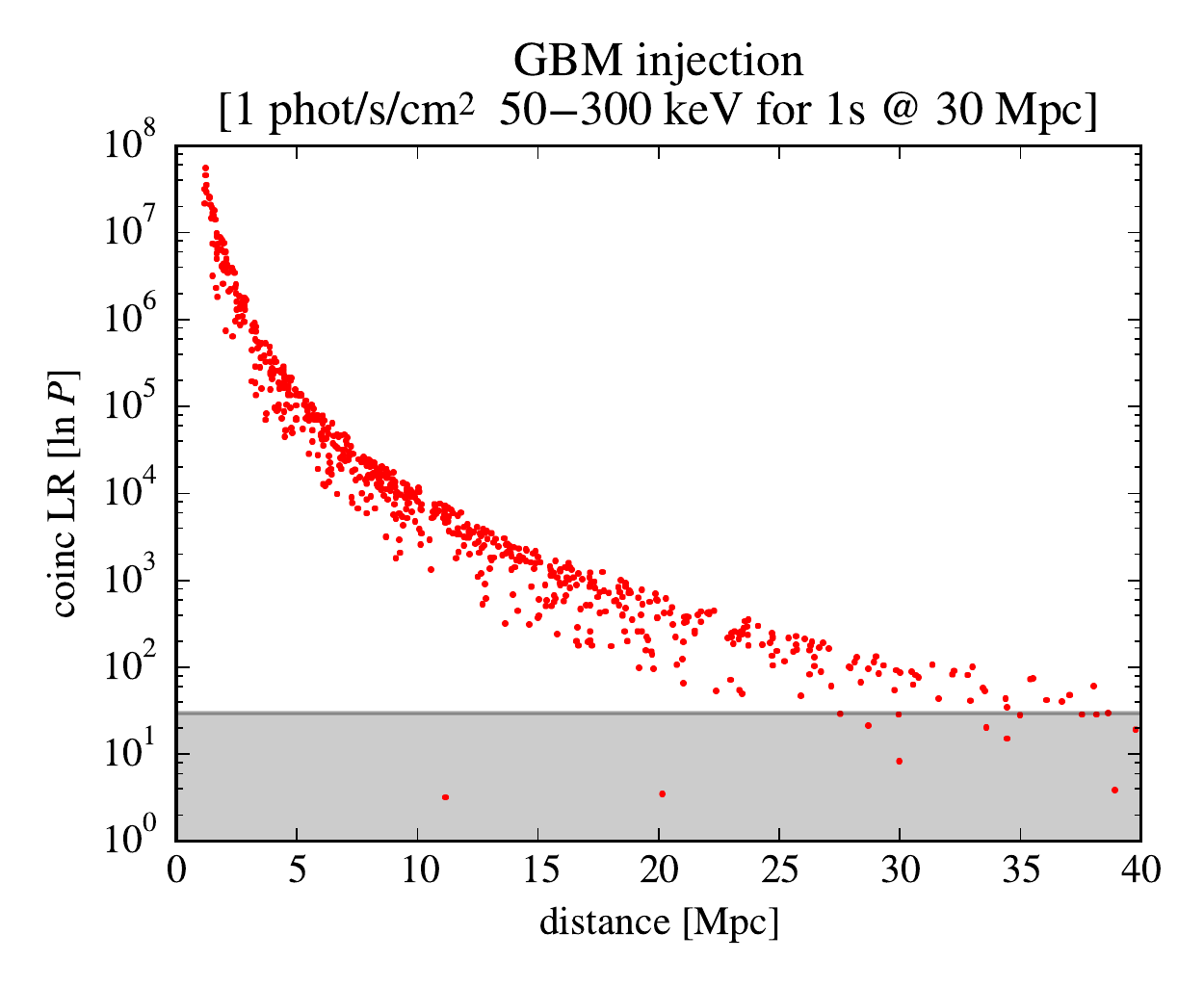}
        \caption{GBM log-likelihood
        ratio after folding in GW sky prior information, as a function of
        injection distance for a standard-candle weak GRB simulation calibrated
        to a flux of 1 photon/s/cm$^2$ in 50--300 keV at 30 Mpc (about
        $6.6\times10^{46}$ erg). The shaded region shows a nominal likelihood-ratio
    threshold of 30.}
\label{fig:gbminjection}
\end{center}\end{figure}

Finally we place a nominal threshold of 30 on the coincident GBM likelihood
ratio, which corresponds roughly to that of the weakest un-triggered GRBs
detectable by this method (figure~\ref{fig:gbminjection}). After this threshold and the previously mentioned
selection cuts, there is only one coincident GW-GBM event remaining, which has
a combined SNR $\rho_c$ of 8.1 (figure \ref{fig:cumhist}), $\sim$26\% lower than the
previous loudest event.

The cumulative histograms of signal-to-noise of the time-shift
GW background events before and after EM coincidence are shown
in figure \ref{fig:cumhist}. This can be qualitatively compared to the
published background distribution shown in figure \ref{fig:ihopedistribution}.

\Needspace*{2\baselineskip}
The original low-threshold GW events used in this analysis were subject
to a single detector SNR threshold of 5.5. This means that for a
double-coincident H1L1 event, the absolute minimum network SNR is
$\sqrt{2\times5.5^2} = 7.78$, before any reduction by the $\chi^2$ consistency
check (equation \ref{eqn:new_snr}). Near this threshold, one must be careful
when assuming simple range scaling relationships between combined SNR
thresholds and effective range of the analysis because of the upstream cuts.

Past the knee in the original GW background distribution $\rho_c \gtrsim 9$,
the inverse relationship between distance and combined SNR should be robust, so
we can estimate a factor of $\gtrsim 15\%$ increase in range for this sample at
equivalent false-alarm rate given events where EM coincidence can provide at
least a factor of 1/1000 background rejection. While the background sample used
in this study reliably probes about three orders of magnitude in rate, the
extended background studies from \cite{Colaboration:2011nz} and represented in
figure \ref{fig:ihopedistribution} show that for H1L1 events near the end of
the LIGO S6 with chirp mass $3.48 \le \mathcal{M}_c < 7.40$, the background
distribution scales roughly as $\sim$$100^{-\rho_c}$ in combined SNR over eight
orders of magnitude in rate.  For the special case in which both a prompt
gamma-ray {\em and} X-ray afterglow counterpart are observable, the accidental
coincidence factors measured here give a large joint rate of background
rejection around $\sim$10$^7$, which implies a factor of $\sim$1.5 increased
range to such events if maintaining a fixed false-alarm-rate given an original
threshold of $\rho_c > 10$ \citep{Camp:2013cwa} before EM coincidence.

\subsection{Efficiency for simulated signals}

\begin{figure} \begin{center}
        \includegraphics[width=3.4in]{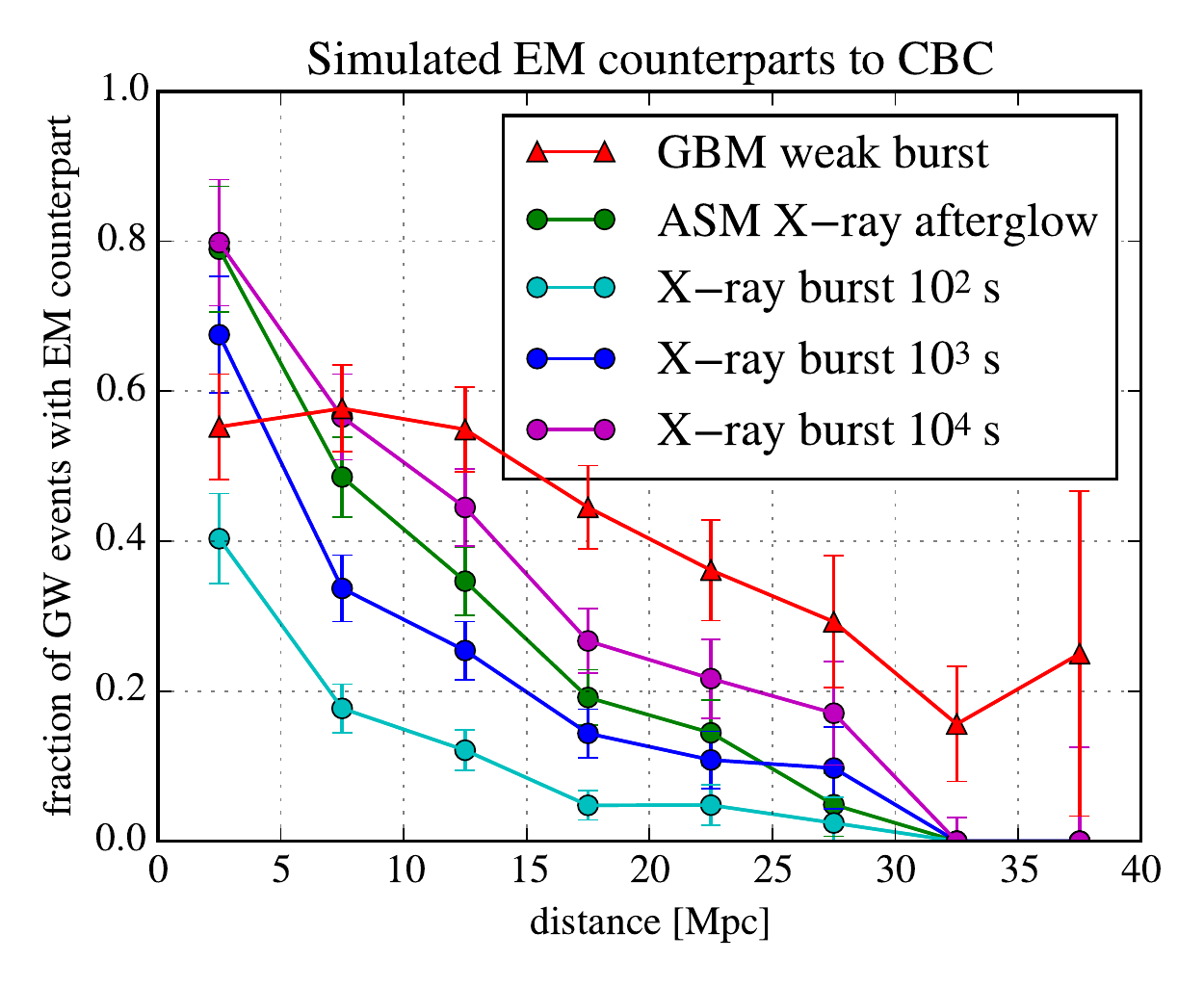} \caption{Fraction of
            EM counterpart injections to simulated GW triggers
            which pass thresholds in the EM analysis, as a function of source
            distance. The EM efficiency is limited by coverage, EM sensitivity,
and the success of GW sky localization. The gamma-ray simulation shown is a weak prompt signal
            lasting 1 second with a standard-candle amplitude of 1
            photon/s/cm$^2$ in 50--300 keV at 30 Mpc, and following a normal
            band spectrum according to table \ref{tab:bandspectra}. This
            corresponds to a total energy release of $6.6\times10^{46}$~erg. The
            ASM X-ray afterglow injection represents a typical X-ray short GRB afterglow shown
            in figure \ref{fig:afterglowlc}, and the X-ray burst injections with varying
            durations are as shown in figure~\ref{fig:afterglowlc}.}
    \label{fig:emeff}
    \end{center}
\end{figure}

The efficiency of the EM pipelines to our standard-candle signal injections is
shown in figure \ref{fig:emeff}. The efficiency fractions are calculated from
the end-to-end simulation of a joint NS/NS inspiral signal with corresponding
prompt and afterglow EM counterpart, and the process is triggered by a
low-threshold ($\rho_c \gtrsim 8.5$) GW candidate. The presence of a GW trigger
is required, and the represented fraction does not include efficiency factors
from the analysis of GW data itself. However, the EM follow-up
efficiency is still generally influenced by the quality of the GW sky localization.

GBM views the entire unocculted sky (65\%) when not in the South Atlantic
Anomaly ($\sim$15\%), and this duty-cycle dominates the efficiency factor out
to 40 Mpc.  This is not surprising as our injeciton amplitude was chosen to be
moderately detectable at 30 Mpc (and still several hundred times weaker than a
typical sGRB).

\Needspace*{2\baselineskip}
The chance of detecting an X-ray afterglow signal with ASM is much more
variable (figure \ref{fig:asm2}) due primarily to the large variability in
delay between onset of the afterglow and the first available measurement. The
ASM follow-up is also more sensitive to the sky localization accuracy from the
GW trigger due to the choice to follow-up only the most probable 200 individual
galaxy host locations. Increased distance both increases the sky area
uncertainty (due to decreased GW SNR), and increases the area-density of
galaxies on the sky. We do not observe ASM counterparts above threshold beyond
$\sim$30 Mpc. For the X-ray burst model waveforms, longer duration bursts (at
equivalent total fluence) were relatively easier to detect as they better
matched the ASM cadence.

\Needspace*{3\baselineskip}
\subsection{Discussion}

In this paper, we introduced a strategy to search for high-energy
EM counterparts to GW binary colescense events in
archival satellite survey data. We designed an end-to-end GW candidate follow-up
pipeline and tested it on a large number of background binary colescence GW
events from time-shifted initial LIGO/Virgo data during their most recent
science runs (S6/VSR2+3). The representative noise events were subject to a
fully-coherent Bayesian parameter estimation analysis in order to obtain sky and
distance posterior probability distributions. These distributions were used to
obtain a set of probable hosts from a catalog of nearby galaxies.

Two custom follow-up methods were designed to search for both a prompt
gamma-ray counterpart in offline data from the Fermi Gamma-ray Burst Monitor
(GBM) within $\pm$30s of the GW coalesence time and consistent with the GW sky
location, as well as a generic family of X-ray afterglow lightcurves in data
from the RXTE All-Sky Monitor (ASM) at the locations of possible host galaxies,
and parametrized by a generic broken power-law with characteristic timescales
of minutes to days. The requirement of a GBM or ASM coincident counterpart reduced
the number of background events by factors of $10^{-4}$ and $10^{-3}$
respectively, reducing the GW ampltiude of the loudest suriving background
event by $\sim$15--20\%.

Both EM pipelines were tested on a set of joint GW-EM simulated signals, where
GW simulations corresponded to NS/NS mergers at random times and sky
locations/orientations. The GBM follow-up pipeline maintained sensitivity to
weak coincident prompt bursts with a normal GRB specturm and a fluence
corresponding to an isotropic-equivalent energy release of
$E_\mathrm{iso}\sim6.6\times10^{46}$ erg over 1s, detecting about 55\% of the
brightest counterparts (due to Earth occultation) and about 20\% of
counterparts placed at 40 Mpc (the most distant simulations due to GW
sensitivity). The simulated signal was several hundred times weaker than a
typical sGRB, implying a wide range of possible GRB luminosities that could be
probed using GW triggers up to advanced LIGO distances. The ASM follow-up
pipeline was able to detect a typical sGRB afterglow signal with about 40\%
efficiency at 10 Mpc.

The strategy of targeted offline follow-up of GW candidates in EM data
presented here is one of a couple possibilities for joint GW-EM observation --
others being rapid online GW analysis to alert pointed telescopes, and using
known EM events like GRB alerts to do a targeted specialized search in GW data.
All of these strategies will likely be in use in the advanced detector era.
The main benefits of having archival EM data at hand is that survey data is
continuously available, and not rate or schedule limited. Thousands or even
millions of GW candidates can be followed-up offline, provided
there is a low-enough probability of accidental coincidence. In the case of a
confident GW detection, automated archival EM searches
will also be able to look for subthreshold/untriggered bursts
\citep{Gruber:2012ny,Blackburn:2013ina}, potentially exotic EM emission
scenarios (e.g. soft flares which are difficult to target on-board due to large
background variation), and in the case of no counterpart, place upper limits on
the EM emission thorough software simulations.

While Fermi will continue operating in the foreseeable future, the RXTE mission
ended in early 2012 after 16 years of operation. In the advanced LIGO era,
the MAXI mission
\citep{2009PASJ...61..999M} on board the ISS has a similar wide-field soft
X-ray camera and continuously surveys the X-ray sky at a sensitivity several
times better than RXTE-ASM. Another promising candidate for X-ray follow-up is ISS\mbox{-}Lobster \citep{Camp:2013cwa} -- a
NASA-proposed wide field ($30^\circ \times 30^\circ$) soft X-ray imaging
telescope which would also be able to survey the sky (as well as do directed
observations). The wide field imager on ISS-Lobster would be $\sim$100 times as
sensitive as RXTE-ASM, matching well the expected factor of 10 increase in GW
horizon distance for advanced LIGO. Moreover the ability of ISS-Lobster to
repoint for an online GW trigger would allow it to begin
observing a rapidly fading afterglow signal as soon as it appears on the right
side of the Earth. Both Fermi-GBM and ISS\mbox{-}Lobster would be able to
search for counterparts with or without the help of a galaxy catalog due to
their extremely wide field-of-view.

\section{Acknowledgements}

We thank Colleen Wilson-Hodge for assistance with the GBM direct response
model. We would also like to thank Collin Capano, Kipp Cannon, Thomas Dent, Leo
Singer, Peter Shawhan, Ruslan Vaulin, Xilong Fan and the LIGO Burst and CBC working groups
for helpful discussion and ideas.  LB did much of this work under the support
of a the NASA Postdoctoral Program fellowship at GSFC, administered by Oak
Ridge Associated Universities through a contract with NASA. J.V. was supported
by the research programme of the Foundation for Fundamental Research on Matter
(FOM), which is partially supported by the Netherlands Organization for
Scientific Research (NWO), and by Science and Technology Facilities Council
(STFC) grant ST/K005014/1. The authors would also like to acknowledge the
support of the NSF through grant PHY-1204371. Finally we thank the Albert
Einstein Institute in Hannover, supported by the Max-Planck-Gesellschaft, for
use of the Atlas high-performance computing cluster.

\bibliography{nasaem}

\end{document}